\documentclass[%
 reprint,
superscriptaddress,
nobibnotes,
 amsmath,amssymb,
 aps,
prx,
]{revtex4-1}

\usepackage{comment}
\usepackage{graphicx}
\usepackage{dcolumn}
\usepackage{bm}

\usepackage{appendix}

\usepackage{graphicx}
\usepackage{dcolumn,amsthm}
\usepackage{bm}

\usepackage{txfonts}
\usepackage{dsfont}
\usepackage{xcolor} 
\usepackage[english]{babel}

\usepackage[colorlinks=true,linkcolor=blue,citecolor=blue,urlcolor=blue]{hyperref}

\newcommand{\ie}{\emph{i.e.,\ }}

\newcommand{\<}{\langle}
\renewcommand{\>}{\rangle}

\providecommand{\tr}{{\rm Tr}}

\renewcommand{\Re}{\mathbb{R}}

\definecolor{royalblue}{HTML}{4169e1}

\begin{document}


\title{Detection of mode-intrinsic quantum entanglement}

\author{Carlos E. Lopetegui}\affiliation{Laboratoire Kastler Brossel, Sorbonne Universit\'{e}, CNRS, ENS-Universit\'{e} PSL,  Coll\`{e}ge de France, 4 place Jussieu, F-75252 Paris, France}
\author{Mathieu Isoard}\affiliation{Laboratoire Kastler Brossel, Sorbonne Universit\'{e}, CNRS, ENS-Universit\'{e} PSL,  Coll\`{e}ge de France, 4 place Jussieu, F-75252 Paris, France}
\author{Nicolas Treps}\affiliation{Laboratoire Kastler Brossel, Sorbonne Universit\'{e}, CNRS, ENS-Universit\'{e} PSL,  Coll\`{e}ge de France, 4 place Jussieu, F-75252 Paris, France}
\author{Mattia Walschaers}
\affiliation{Laboratoire Kastler Brossel, Sorbonne Universit\'{e}, CNRS, ENS-Universit\'{e} PSL,  Coll\`{e}ge de France, 4 place Jussieu, F-75252 Paris, France}




\date{\today}

\begin{abstract}
Quantum correlations are at the core of the power of quantum information and are necessary to reach a quantum computational advantage. 
In the context of continuous-variable quantum systems,
another necessary ressource for quantum advantages is non-Gaussianity. 
In this work, we propose a witness, based on previously known relations between metrological power and quantum correlations, to detect a strong form of entanglement that only non-Gaussian states possess and that  
cannot be undone by passive optical operations, i.e., entanglement in all mode bases. The strength of our witness is two-fold: it only requires measurements in one basis to check entanglement in any arbitrary mode basis; it can be made applicable experimentally using homodyne measurements and without requiring a full tomography of the state. 
\end{abstract}

\maketitle


\section{Introduction}\label{sec:intro}
Major efforts are being conducted worldwide to harness the power of quantum mechanics for technological applications. The use of these systems for performing computations beyond the capabilities of classical processors stands as one of the most encouraging perspectives. A wide variety of physical platforms are being considered for this purpose \cite{IBM_superconduncting,GoogleQ,Henriet2020,Lukin2023,ion-traps,Alice_and_Bob_architecture,Psiquantum2023,Bourassa2021}, among which optical systems prove to be a promising framework given its potential for scalability \cite{Larsen2019, Asavanant2019}. 

The full power of optical quantum technologies is attained when we can coherently control the collective state of many modes. These collective states manifest entanglement, a unique feature of quantum systems that has challenged the common conceptions of physicists over the last century \cite{EPR1935,Schrod1935,Horodecki2009}. Entanglement and quantum correlations in general, reflect in measurement statistics that are not consistent with classical correlations.  Mathematically this implies that they cannot be explained using a local hidden variable model \cite{Bell1964}.
In the context of quantum information processing and quantum computing entanglement has revealed to be an essential resource \cite{Acin2018, Jozsa2003}. However, in continuous variables (CV) systems, entanglement is not enough if the state is fully Gaussian, or if its associated Wigner function -- which describes all the statistical properties of the quantum state -- is positive. Indeed, it has been shown that negativity in the Wigner function is a necessary resource for realizing sampling protocols that cannot be efficiently simulated with classical resources \cite{Mari_Eisert_2012}. This result significantly boosted the already ongoing efforts aimed at producing non-Gaussian states \cite{Ra2020, Endo2023, Neergaard-Nielsen2006, Wakui2007}, with remarkable results achieved in microwave cavities and superconducting circuits in the production of exotic non-Gaussian states \cite{Kudra2022,Touzard2018}. 

Yet, recent results have pointed out that not any kind of quantum entanglement (even if the state is Wigner negative) is sufficient to render computations with CV quantum systems hard to simulate with a classical device. In particular, in Ref. \cite{Ulysse_Mattia_2022} it was shown that in a certain family of sampling protocols a strong form of quantum entanglement is required: not passive separability, i.e., the fact that entanglement cannot be undone with optical passive transformations (beamsplitters and phase shifters). This kind of entanglement was previously introduced in \cite{Sperling2019}, from a different perspective. They refer to it as mode-independent quantum entanglement and point out its potential usefulness for communication protocols. We refer to this kind of quantum correlations as \textit{mode-intrinsic entanglement} \footnote{We prefer this terminology over \textit{mode-independent} quantum entanglement, used in \cite{Sperling2019}, because we consider the latter could mislead the reader to think that the amount of entanglement in all basis is the same, which is generally not the case.}. Finding and characterizing non-Gaussian states that exhibit this refined quantum feature and that can be produced in the lab are thus important steps in the domain of CV quantum information. 
 
\par
Many witnesses and criteria have been derived for detecting entanglement in Gaussian states \cite{Simon2000, Duan2000, Giovannetti2003, VanLoock2003, Abiuso2021}. All of them rely in some way in the knowledge of the covariance matrix. Once measured in one basis, such a matrix can be obtained in any mode basis by means of an orthogonal transformation. For this reason these methods are quite versatile for studying entanglement in every mode basis. However, Williamson’s decomposition \cite{williamson1936,PhysRevA.71.055801} implies that there always exists a basis in which the covariance matrix is consistent with a separable Gaussian state, independently of the nature of the quantum state itself. Therefore, these methods are not suitable for demonstrating mode-intrinsic entanglement. Furthermore, it shows that mode-intrinsic entanglement is inherently a non-Gaussian feature.


 Over the years many protocols have been proposed to detect entanglement in non-Gaussian states. If one has access to the density matrix of a given state it is possible to use the positive partial transpose (PPT) criterion \cite{Peres1996} to detect quantum correlations. However, the reconstruction of the density matrix requires a full tomography of the state \cite{Lvovsky2009}, a very demanding and time-consuming experimental task. Efforts to circumvent this issue have lead to the introduction of different methods to witness quantum correlations \cite{Vogel2005,Rodo2008,Walborn2011,Saboia2011,Gessner2016,EPR_Metr2021,Garttner2023,Xiaoting_2024, valido2014}.\par 
Among those there is a subset \cite{Gessner2016,EPR_Metr2021} that rely on connections to quantum metrology \cite{Giovannetti2011, Pezze2018, Toth2014}. 
 These witnesses are based on the Fisher Information \cite{Liu_2020}, a quantity that reflects the highest sensitivity with which we can estimate a parameter of interest using measurements of an observable. These methods rely on the idea that quantum correlations can lead to metrological sensitivities that could not be achieved with only classical correlations. CV-specialized protocols based on these witnesses \cite{steering_2022,David_2023} have been shown to succeed in detecting quantum correlations in relevant non-Gaussian states where usual Gaussian criteria fail. This suggests that non-Gaussian entanglement is not only crucial in the framework of quantum computation but can also potentially offer advantages in sensing applications. This idea relates naturally to recent results \cite{Halpern_2022} that link the negativity of some quasiprobability distributions to enhancements in quantum phase estimation in quantum optics.\par
All the methods we have mentioned so far to detect entanglement in non-Gaussian states are developed to detect entanglement once we have fixed a mode basis of interest. On the other hand, to study entanglement in every mode basis, a method that retains the versatility of the Gaussian protocols, but potentially succeeds where the former fails, is necessary. In this paper, we propose such a protocol based on metrological witnesses. We show that our protocol works for relevant non-Gaussian states that can be produced with state-of-the-art experiments, namely photon subtracted states \cite{Ra2020, Endo2023}. Likewise, our witness also demonstrates effectiveness in multimode systems such as (non-Gaussian) cluster states -- that can be obtained by de-gaussifying readily available large CV cluster states \cite{Larsen2019, Asavanant2019,Ra2020}, which have been widely studied for their scalability in the pursuit of large-scale quantum computing. While sensitive to losses, the mode-intrinsic-entanglement witness that we suggest here can be applied experimentally through homodyne detection and the estimation of the corresponding classical Fisher information -- using the Hellinger distance as proposed in Ref. \cite{David_2023}; it thus avoids full tomography of the state, making metrological witnesses a very efficient tool in the detection of mode-intrinsic entanglement and non-Gaussian features. Moreover, it proves useful for the less specific task of studying the entanglement properties of a given state beyond the basis in which measurements are performed, with higher accuracy than Gaussian methods. \par
The paper is organized as follows: in Sec.~\ref{sec:Entanglement_and_FI} we introduce the metrological entanglement witness derived in Ref. \cite{Gessner2016} and extend it to the detection of mode-intrinsic entanglement in Sec.~\ref{sec:witness-passive-separability}. Then, in Sec. \ref{sec:application_witness}, we apply our witness to experimentally relevant multimode states, namely photon subtracted states. In the last section (Sec.~\ref{sec:relaxation_criteria}), we consider a lower bound of the mode-intrinsic entanglement witness derived in the previous sections; we show that this witness is experimentally accessible through homodyne measurements; in particular, we test this lower bound witness on a two-photon subtracted state using simulated experimental data.

\section{Metrology based entanglement witnesses}\label{sec:Entanglement_and_FI}

\par 
In this section we introduce metrological witnesses of entanglement in which the protocol that we present is based \cite{Gessner2016}. These witnesses rely on the analysis of a specific metrological task. We consider two parties, Alice and Bob, who share a state $\hat{\rho}$. 

\begin{figure}
\includegraphics[width =0.8\linewidth]{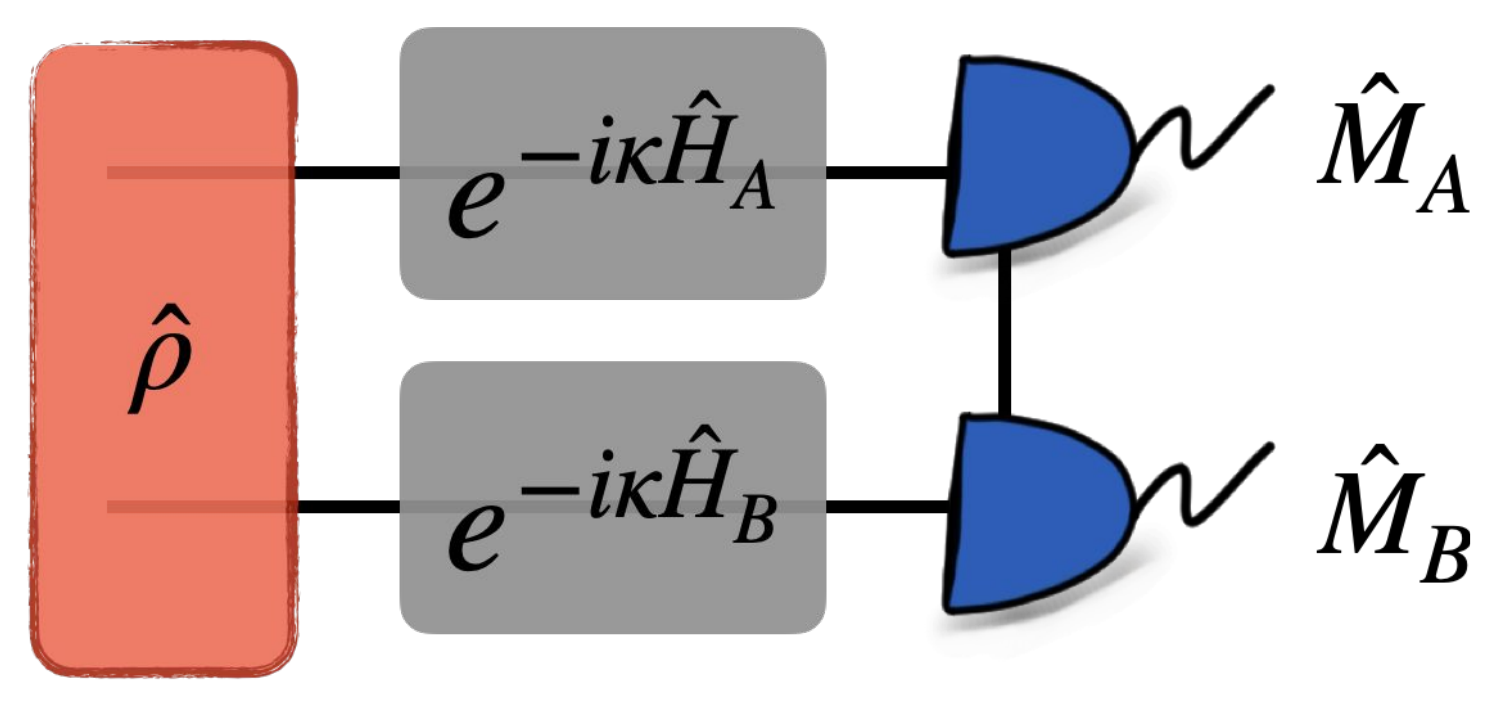}
    \caption{ Schematic of the metrological protocol in which the entanglement witness that we present is based on. Alice and Bob share a state $\hat \rho$ and try to estimate a parameter $\kappa$ encoded in a unitary evolution under the hamiltonian $\hat H_A+\hat H_B$, given by a sum of local Hamiltonians, which implies that the evolution can actually be factorized. To perform the estimation they can perform any measurements $\hat M_A(B)$. The line joining the measurement devices is made to highlight that they can perform any set of non-local compatible measurements. }
    \label{fig:metrological_protocol}
\end{figure}

They are interested in jointly estimating a parameter $\kappa$ that they encode through a unitary transformation induced by the sum of local generators $ \hat H = \hat{H}_A + \hat{H}_B$ (see Fig.\ref{fig:metrological_protocol}). In this case, the density matrix of the state transforms as $\hat \rho_\kappa = \exp(-i \kappa  \hat H) \, \hat \rho \, \exp(i \kappa \hat H)$. The sensitivity with which they can estimate the parameter, given that they can measure some (possibly non local) observables $\hat{M}_{A(B)}$, is upper bounded by the Fisher information associated to the probability distribution of outcomes of these observables.
The Fisher information is given by 
\begin{equation}\label{eq:FI-definition}
\begin{split}
F(p(M_{A}, M_{B}|\kappa))=\int_{ M_A, M_B} \! \! \!  \! \! \! d M_A \, d M_B & \left(\frac{\partial \mathcal{L}(p(M_A, M_B|\kappa))}{\partial \kappa}\right)^2 \\
& \times p(M_A, M_B|\kappa),     
\end{split}
\end{equation}
where $\mathcal{L}(p(M_A, M_B|\kappa))=\log(p(M_A, M_B|\kappa))$, stands for the logarithmic likelihood, and $p(M_A, M_B|\kappa) = \tr\left( \hat \rho_{\kappa} \, \hat \Pi_{AB} \right) $ is the joint probability distribution conditioned by the parameter $\kappa$, where $ \hat \Pi_{AB} = |M_A, M_B \rangle \langle M_A, M_B|$ is the projector into the outcomes $(M_A, M_B)$ of $\hat M_A \otimes \hat M_B$. 
The ultimate sensitivity achievable in the estimation of $\kappa$ is given by the quantum Fisher information (QFI) of the state, defined as the maximum Fisher information, optimized over all possible measurements, \ie 
\begin{equation} \label{eq:comp_FI_QFI}
    F_{Q}(\hat{\rho}, \hat H)=\max_{\hat M_A, \hat M_B} F\left[ \tr\left( \hat \rho_{\kappa} \, \hat \Pi_{AB} \right)\right].
\end{equation}
It can be shown that for pure states 
\begin{equation}
    F_{Q}\left(| \psi \rangle\langle \psi|, \, \hat H \right)=4 \text{Var}\left(| \psi\rangle \langle \psi|,\hat{H} \right),
\end{equation}
where $\text{Var}\left(| \psi\rangle \langle \psi|,\hat{H} \right)$ is the variance of the generator $\hat H$ in the state $|\psi \rangle$. In the case of mixed states there is also an analytical expression for the quantum Fisher information, but for the moment it suffices to say that $F_{Q}(\hat{\rho},\hat H)\leq 4 \text{Var}(\hat \rho, \hat H)$.
Another useful property for our purpose is the aditivity of the QFI, namely 
\begin{equation}
    F_{Q}(\hat{\rho}_{A}\otimes \hat{\rho}_B, \hat{H}_A+\hat{H}_B)=F_{Q}(\hat{\rho}_{A},\hat{H}_A)+F_{Q}( \hat{\rho}_B,\hat{H}_B),
\end{equation}
where $\hat{\rho}_{A(B)} = \tr_{B(A)}(\hat \rho)$ is the reduced density matrix.
Combining these two properties and the convexity of the quantum Fisher information, one can show that that if Alice and Bob share a separable state $\hat{\rho}$ it leads to the inequality \cite{Gessner2016}
\begin{equation}\label{eq:basic-inequality}
    F_{Q}(\hat\rho,\hat H =  \hat{H}_A+\hat{H}_B)\leq 4 \left(\text{Var}(\hat{\rho}_A,\hat{H}_A)+\text{Var}(\hat{\rho}_B,\hat{H}_B) \right).
\end{equation}
This inequality can only be violated by entangled states and can be thus used as a witness.
In practical terms this implies that if they share an entangled state they better join forces than trying to estimate independently the parameter. One would be tempted to say that this implies that entanglement offers a metrological advantage in this setup; yet the situation is a bit more subtle than that, because we are comparing the estimation they can do jointly with respect to what they can estimate with their noisy reduced states.
However, this does not imply that there is no better pure single mode state they could independently use and get each a better Fisher information. \par
The full power of this kind of inequalities can be exploited when we optimize over a set of allowed choices of generators. For any $\hat{H}_{A(B)}$, the violation of the metrological inequality provides a sufficient condition for entanglement. Yet, if a violation is not attained it might easily be that there is a different choice of generators that produces one. If Alice and Bob each have access to a set of local generators 
$\mathcal{A}_{A(B)}=\{\hat{H}_{A(B)}^{(1)},...,\hat{H}_{A(B)}^{(N_{A(B)})}\}$,
 where $N_{A(B)}$ is the number of generators in each set,
they can choose to implement generators given by $\hat{H}_{A(B)}=\mathbf{c}_{A(B)}^T\mathcal{A}_{A(B)}$. We group $\mathbf{c}_{A(B)}$ in a single vector $\mathbf{c}=\{\mathbf{c}_{A},\mathbf{c}_{B}\}$, and similarly $\mathcal{A} = \{\mathcal{A}_{A}, \mathcal{A}_{B}\}$. This allows to write the Quantum Fisher Information in a matrix form as \cite{Gessner2016}
\begin{equation}\label{eq:QFI_from_matrix}
F_{Q}(\hat{\rho},\mathbf c^T \mathcal{A})=\mathbf{c}^T \mathcal{Q}_{\hat \rho}^{\mathcal{A}}\mathbf{c},
\end{equation}
where $\mathcal{Q}_{\hat\rho}^{\mathcal{A}} $ is the Quantum Fisher Information matrix and its elements are given, provided the spectral decomposition $\hat \rho=\sum_{k}p_k |\psi_k\>\<\psi_k|$, by 
\begin{equation} \label{eq:QFI_from_matrix-coeffs}
\left(\mathcal{Q}_{\hat{\rho}}^{\mathcal A}\right)_{i,j}^{(m,n)}=\sum_{k,l} \frac{(p_k-p_l)^2}{p_k+p_l} \<\psi_k|\hat{H}_{i}^{(m)}|\psi_l\>\<\psi_l|\hat{H}_{j}^{(n)}|\psi_k\>,
\end{equation}
where the sum extends over all pairs $k,l$ such that $p_k+p_l\neq 0$. The state $\hat\rho$ is typically considered to be the one after the implementation of the evolution. Yet here we consider only small values of the parameters, so that the spectral decomposition considered is that of the probe state, and all higher orders on the value of the parameters are omited. \par
The right hand side of inequality \eqref{eq:basic-inequality} on the other hand corresponds to the variance of the generator in the product state of the local reduced states, \ie $\Pi(\hat \rho)=\hat \rho_A \otimes \hat\rho_B$. This term can be expressed also in a matrix form by means of the covariance matrix of the generators defined by 
\begin{equation} \label{eq:var-matrix-coeffs}
\Gamma_{i,j}^{(m,n)}=Cov(\hat{H}_{i}^{(m)},\hat{H}_{j}^{(n)})_{\Pi(\hat\rho)}.
\end{equation}
The variance of the generator for a particular choice of $\mathbf{c}$ is given by 
\begin{equation}\label{eq:var-matrix-form}
\text{Var}(\mathbf{c}_{A}^{\rm \scriptscriptstyle T} \mathcal{A}_{A}) + \text{Var}(\mathbf{c}_{B}^{\rm \scriptscriptstyle T} \mathcal{A}_{B})=\mathbf{c}^T \Gamma_{\Pi(\hat\rho)}^{\mathcal{A}}\mathbf{c}. 
\end{equation}
Combining \eqref{eq:QFI_from_matrix} and \eqref{eq:var-matrix-form} into \eqref{eq:basic-inequality} we get that, for any separable state  
\begin{equation} \label{eq:W-witness}
\mathbf{c}^T (\mathcal{Q}_{\hat \rho}^{\mathcal A}-4 \Gamma_{\Pi(\hat \rho)}^{\mathcal A})\mathbf c\leq 0, \qquad \forall \mathbf{c}.
\end{equation}
Given that this condition is satisfied for all choices of $\mathbf{c}$, it can be rewritten as 
\begin{equation}\label{eq:eigenv_ineq}
 \lambda_{\rm \scriptscriptstyle max}\left(\mathcal{Q}_{\hat \rho}^{\mathcal A}-4 \Gamma_{\Pi(\hat \rho)}^{\mathcal A}\right)\leq 0,
\end{equation}
where $\lambda_{\rm \scriptscriptstyle max}(M)$ stands for the maximum eigenvalue of $M$. This inequality provides us a witness of entanglement 
\begin{equation}\label{eq:entanglement_witness_matrix_form}
    E_{\rho}^{\mathcal A} \equiv \lambda_{\rm \scriptscriptstyle max}\left(\mathcal{Q}_{\hat \rho}^{\mathcal A}-4 \Gamma_{\Pi(\hat \rho)}^{\mathcal A}\right).
\end{equation}
If this witness takes possitive values we can certify the presence of entanglement. Notably, if the set of generators is a complete set and the state is a pure state the latter becomes not just a sufficient but also a necessary condition \cite{Gessner2016}. \par
All the above results can be straightforwardly generalized to multimode systems and to multipartite entanglement detection. In this case, the $m$ parties have access to a set of local generators $\mathcal{A}_m = \{ \hat H^{(1)}_m, \ldots, \hat H^{(N_m)}_m \}$. The elements of the Quantum Fisher Information matrix and the covariance matrix can be then computed from equations \eqref{eq:QFI_from_matrix-coeffs} and \eqref{eq:var-matrix-coeffs}, with $\Pi(\hat \rho)=\hat \rho_{\scriptscriptstyle \mathcal{P}_1} \otimes \ldots \otimes \hat\rho_{ \scriptscriptstyle \mathcal{P}_p}$, where we split the $m$-multimode system into $p$ partitions $\mathcal{P}_i$ to detect $p$-partite entanglement. 


\section{Witness mode-intrinsic entanglement} \label{sec:witness-passive-separability}

A key aspect of non-Gaussian entanglement lies in its robustness against mode basis change. Contrary to Gaussian states whose entanglement can always be destroyed through a passive linear optics transformation, no such transformation can entirely remove the entanglement present in some non-Gaussian states; these states are said to be \textit{not passively separable} and we call this type of entanglement \textit{mode-intrinsic entanglement}. To detect such entanglement we propose to use the metrological witness \eqref{eq:eigenv_ineq} introduced in Sec.\ref{sec:Entanglement_and_FI} as follows: a state will be mode-intrinsic entangled if in every mode basis $ E_{\rho}^{\mathcal A} > 0$.

In the following we focus on mutlimode quantum optical systems \cite{Treps2020}. 
In CV quantum optics the natural and fundamental observables are the amplitude and phase quadratures $\hat q_k$ and $\hat p_k$, which are related to creation and annihilation operators of the corresponding bosonic field by  
\begin{equation}
    \hat a_k = \frac{\hat q_k + i \hat p_k}{2}, \quad  \hat a_k^\dagger = \frac{\hat q_k - i \hat p_k}{2}, \quad \forall k \in \{1, \ldots, m\},
\end{equation}
where  $\hat a_k$ and $\hat a^{\dagger}_k$ are, respectively, the creation and annihilation operators corresponding to mode $k^{th}$ out of $m$ available modes. Quadrature operators satisfy the canonical commutation relation $[ \hat q_k, \hat p_l] = 2 i \, \delta_{k,l}$. The statistics of these operators can be recovered experimentally through homodyne measurements \cite{Leonhardt1995}, making them a judicious choice to construct a set of generators that the $m$ parties involved in the multimode system can experimentally implement. 
To that aim, we introduce the following generators of order $N$
\begin{equation} \label{eq:generators-quadratures}
    \hat{H}_i^{\mathcal{S}} = \mathcal{S}\left( \hat q_1^{k_1^{(i)}} \ldots \hat q_m^{k_m^{(i)}} \, \hat p_1^{k_{m+1}^{(i)}} \ldots \hat p_m^{k_{2m}^{(i)}} \right), \, \text{with}\, \sum_{j=1}^{2m} k_j^{(i)} = N,
\end{equation}
where $\mathcal{S}( \hat H)$ is the symmetrised version of $\hat H$ such that $\hat{H}_i^{\mathcal{S}}$ is Hermitian.
As one can see from the above expression, $\hat{H}_i^{\mathcal{S}}$ contains products of $N$ quadrature operators and is entirely described by the combination $(k_1^{(i)}, \ldots, k_{2m}^{(i)})$, where each $k_j^{(i)}$ corresponds to the power of the quadrature operators in the decomposition \eqref{eq:generators-quadratures} and whose sum equals $N$. Likewise, this type of generators can be \textit{non-local}. Here, locality is defined as follows: \textit{local} generators only belong to one subsystem, while \textit{non-local} generators can be shared between multiple subsystems; for instance the quadrature operator $\hat p_{1}^2$ is a local generator, while $\hat p_{1} \, \hat p_{2}$ is a non-local generator shared between subsystems 1 and 2. Since the metrological protocol described in the previous section is only based on local generators, it may seem unnecessary to include non-local generators. However, they will play a crucial role to correctly implement the mode basis change as we will see below.

Then, we denote by $\mathcal{A}^{\rm \scriptscriptstyle (NL)}_{\scriptscriptstyle N, \scriptscriptstyle m} = \{\hat{H}_1^{\mathcal{S}}, \ldots \hat{H}_{\ell(N,m)}^{\mathcal{S}}\}$ the set of all (local and non-local) symmetrised generators of the form \eqref{eq:generators-quadratures} up to order $N$, in a $m$-mode system.  The superscript (NL) stands for \textit{non-local} and
\begin{equation} \label{eq:length_set}
\ell(N,m)  = 
\sum_{k=1}^N \begin{pmatrix}
    2m+k-1\\
    k
\end{pmatrix}
\end{equation}
is the length of the set. As an illustration, for $m=2$ and $N = 2$ this set of operators reads
\begin{equation} \label{eq:set-generators-order-2}
\begin{split}
     \mathcal{A}^{\rm \scriptscriptstyle (NL)}_{2,2} = &  
 \left\{\hat q_1, \hat p_1, \hat q_2, \hat p_2,\hat q_1^2, \frac{1}{2}( \hat q_1 \hat p_1 + \hat p_1 \hat q_1),\hat p_1^2, \hat q_2^2, \right. \\
& \left. \frac{1}{2}(\hat q_2 \hat p_2 + \hat p_2 \hat q_2),\hat p_2^2, \hat q_1 \hat q_2, \hat p_1 \hat p_2, \hat q_1 \hat p_2, \hat p_1 \hat q_2 \right\}.
\end{split}
\end{equation}
 Following the above notation, we also denote by $\mathcal{A}^{\rm \scriptscriptstyle (L)}_{\scriptscriptstyle N,m}$ the set of symmetrised generators of order $N$ containing only local generators, in an $m$-mode system. In the case of the set \eqref{eq:set-generators-order-2}, this amounts to remove the four generators $\hat q_1 \hat q_2, \hat p_1 \hat p_2, \hat q_1 \hat p_2$ and $\hat p_1 \hat q_2$.

Now that we have introduced the set of local and non-local generators ordered by $N$, we can use them to test mode-intrinsic entanglement of a given quantum state of light. To proceed, we probe entanglement using the metrological witness in every mode bases. It is worth stressing here that it is not necessary to measure the quantum Fisher information matrix $\mathcal{Q}_{\hat \rho}^{\mathcal A}$ as well as the covariance matrix of the local reduced state $\Gamma_{\Pi(\hat \rho)}^{\mathcal A}$ in all bases. Indeed, once these matrices are computed or measured in one basis, one can calculate them in another basis in post-processing. To illustrate this important point, let us consider the unitary matrix
$U(\boldsymbol{\vartheta})$ parameterized by some parameters contained in a vector $\boldsymbol{\vartheta}$
and which enables to go from one set of generators $\mathcal{A}_{\scriptscriptstyle N,m}^{\rm \scriptscriptstyle (NL)}$ in a given basis to another set of generators $(\mathcal{A}_{\scriptscriptstyle N,m}^{\rm \scriptscriptstyle (NL)})^\prime$ in another basis, i.e., $(\mathcal{A}_{\scriptscriptstyle N,m}^{\rm \scriptscriptstyle (NL)})^\prime = U(\boldsymbol{\vartheta}) \,  \mathcal{A}_{\scriptscriptstyle N,m}^{\rm \scriptscriptstyle (NL)}$. The exact form of $U(\boldsymbol{\vartheta})$ will be detailed below after Eq. \eqref{eq:alpha-beta-mode-basis-change}. 
At the level of the quantum Fisher information matrix
the change of basis gives
\begin{equation}\label{eq:transform_QFI}
    \left( \mathcal{Q}_{\hat \rho}^{\mathcal{A}_{\scriptscriptstyle N,m}^{\rm \scriptscriptstyle (NL)}} \right)^\prime = U(\boldsymbol{\vartheta}) \, \mathcal{Q}_{\hat \rho}^{\mathcal{A}_{\scriptscriptstyle N,m}^{\rm \scriptscriptstyle (NL)}} U^T(\boldsymbol{\vartheta}).
\end{equation}
 Therefore, knowing how the non-local set of generators transforms under a mode basis change allows to find the transformation at the level of the matrix $\mathcal{Q}_{\hat\rho}^{\mathcal{A}}$ (and similarly for $\Gamma_{\Pi(\hat\rho)}^{\mathcal{A}} $) used in 
Eq. \eqref{eq:W-witness}, and thus allows to find the metrological witness in the new basis. 

There are some subtleties though: the basis change must include non-local operators to be correct. Indeed, let us take a simple example to illustrate this important point: the generator $\hat p_1^2$ for a two-mode subsystem will be mapped onto $(\cos\theta \hat p_1^{\prime} + \sin \theta \hat p_2^{\prime})^2$ when applying a basis change, and it clearly contains the non-local term $\hat  p_1^{\prime} \, \hat p_2^{\prime}$. Therefore, we need to keep such a non-local term to ensure that matrix $\mathcal{Q}_{\hat\rho}^{\mathcal{A}}$ in Eq. \eqref{eq:transform_QFI} correctly transforms under the mode basis change $U(\boldsymbol{\vartheta})$.

\begin{figure}
\includegraphics[width =\linewidth]{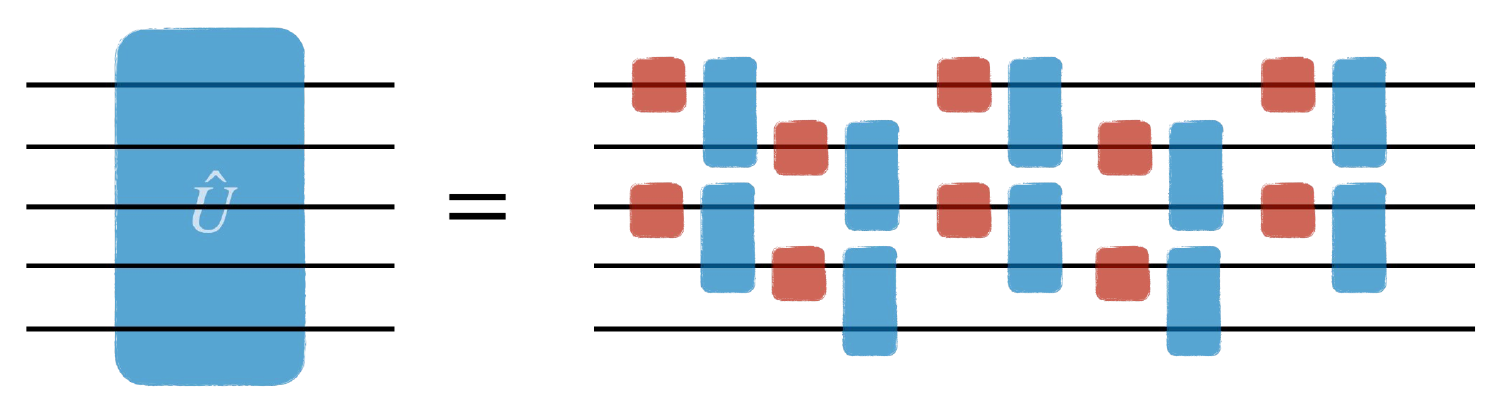}
    \caption{Clements' design for the representation of an arbitrary m-mode passive interferometer \cite{Clements2017}. Red blocks represent local transformations parametrized by an angle $\phi\in \left[0,2\pi\right)$, and blue blocks involving two neighbouring modes represent real beam splitters, parametrized by an angle $\theta \in \left[0,2\pi\right)$. A set of $m$ local rotations was omitted after the last set of beam splitters as they do not modify the entanglement properties of the probe state.  }
    \label{fig:Clements_decomposition}
\end{figure}

However, once each matrix $\mathcal{Q}_{\hat \rho}^{\mathcal{A}_{\scriptscriptstyle N,m}^{\rm \scriptscriptstyle (NL)}}$ and $\Gamma_{\Pi(\hat \rho)}^{\mathcal{A}_{\scriptscriptstyle N,m}^{\rm \scriptscriptstyle (NL)}}$ in Eq. \eqref{eq:W-witness} have been found in the new basis, one should then drop all the rows and columns of these matrices associated to non-local generators. Indeed, the metrological protocol as described in the previous section consists of taking one \textit{local} generator for each party of the multimode state, thus excluding non-local generators shared between subsystems. Keeping this important point in mind, the overall procedure then reads as follows:
\begin{enumerate}
    \item Compute the quantum Fisher information matrix $\mathcal{Q}_{\hat \rho}^{\mathcal A_{\scriptscriptstyle N,m}^{\rm \scriptscriptstyle (NL)}}$ for the whole set of operators $\mathcal{A}_{\scriptscriptstyle N,m}^{\rm \scriptscriptstyle (NL)}$. When the state is pure this matrix equals four times the extended covariance matrix $\Gamma_{\hat \rho}^{\mathcal{A}_{\scriptscriptstyle N,m}^{\rm \scriptscriptstyle (NL)}}$ (whose definition has been given in Eq. \eqref{eq:var-matrix-coeffs} with $\Pi(\hat \rho)$ replaced by $\hat \rho$). 
    
    Then, compute $\Gamma_{\hat \rho}^{\mathcal{A}_{\scriptscriptstyle N,m}^{\rm \scriptscriptstyle (NL)}}$. Both matrices have dimension $\ell(N,m)\times \ell(N,m)$. 
    
    \item Make the change of basis according to Eq. \eqref{eq:transform_QFI} to obtain $ \left(\mathcal{Q}_{\hat \rho}^{\mathcal A_{\scriptscriptstyle N,m}^{\rm \scriptscriptstyle (NL)}} \right)^\prime$ and $\left(\Gamma_{\hat \rho}^{ \mathcal{A}_{\scriptscriptstyle N,m}^{\rm \scriptscriptstyle (NL)}} \right)^\prime$.
    \item Remove the rows and columns related to non-local operators. 
    The final matrices denoted by $\left(\mathcal{Q}_{\hat \rho}^{\mathcal A_{\scriptscriptstyle N,m}^{\rm \scriptscriptstyle (L)}}\right)^\prime$ and $\left( \Gamma_{\hat \rho}^{ \mathcal{A}_{\scriptscriptstyle N,m}^{\rm \scriptscriptstyle (L)} } \right)^\prime$ will have dimension $\ell_{\rm \scriptscriptstyle loc}(N,m)\times \ell_{\rm \scriptscriptstyle loc}(N,m)$, with  $ \ell_{\rm \scriptscriptstyle loc}(N,m) = m \, N \,(N+3)/2 $.
       \item Compute $\left( \Gamma_{\Pi(\hat \rho)}^{\mathcal A_{\scriptscriptstyle N}^{\rm \scriptscriptstyle (L)}} \right)^\prime$. This computation amounts to keep the elements of $\left( \Gamma_{\hat \rho}^{\mathcal A_{\scriptscriptstyle N}^{\rm \scriptscriptstyle (L)}} \right)^\prime$ which only involve two generators $\hat{H}_i^{\mathcal{S}}$ and $\hat{H}_l^{\mathcal{S}}$ belonging to the same subsystem and put zero elsewhere.
       \item Compute $E_{\rho}^{\mathcal A_{\scriptscriptstyle N,m}}(\boldsymbol{\vartheta}) = \lambda_{\rm \scriptscriptstyle max}\left[ \left(\mathcal{Q}_{\hat \rho}^{\mathcal A_{\scriptscriptstyle N,m}^{\rm \scriptscriptstyle (L)}} \right)^\prime - 4 \, \left(\Gamma_{\Pi(\hat \rho)}^{\mathcal A_{\scriptscriptstyle N,m}^{\rm \scriptscriptstyle (L)}} \right)^\prime \right] $, i.e., the entanglement criterion in the new basis. 
       Then, if 
       \begin{equation}\label{eq:PS-metrological-witness}
          \mathcal{W}_Q(\hat \rho, \mathcal{A}_{N,m}) \equiv  \underset{\boldsymbol{\vartheta}}{\text{min}} \,E_{\rho}^{\mathcal A_{\scriptscriptstyle N,m}}(\boldsymbol{\vartheta})   >0,
       \end{equation}
       the state will be not passively separable.
\end{enumerate}
To apply the above procedure we now need to derive the matrix transformation $U(\boldsymbol{\vartheta})$. We recall that a generator $\hat{H}_i^{\mathcal{S}}$ of type \eqref{eq:generators-quadratures} is described by the combination $(k_1^{(i)}, \ldots, k_{2m}^{(i)})$, where each $k_j^{(i)}$ corresponds to the power of each quadrature operator involved in the decomposition \eqref{eq:generators-quadratures}. 
The operators in the new basis are obtained from the operators in the old basis as
\begin{equation}
\left(\hat{H}_i^{\mathcal{S}} \right)^{\prime} =   \sum_l U_{il}(\boldsymbol{\vartheta}) \, \hat{H}_l^{\mathcal{S}}, \quad \forall i \in \{1, \ldots, \ell(N, m)\}
\end{equation}
where
\begin{equation} \label{eq:basis-change-Utheta}
 U_{il}(\boldsymbol{\vartheta}) =   \frac{ 1  }{   \prod_{j=1}^{2m} k_j^{(l)}! } \sum_{\sigma \in S_N}\prod_{s=1}^{N} O_{\alpha_s \,  \beta_{\sigma(s)}}^{\rm \scriptscriptstyle T}(\boldsymbol{\vartheta}),
\end{equation}
with
\begin{equation} \label{eq:alpha-beta-mode-basis-change}
\begin{split}
      \alpha  & = \{\underbrace{1,\ldots 1}_{k_1^{(l)}}, \ldots, \underbrace{2m, \ldots, 2m}_{k_{2m}^{(l)}} \},\\
\beta & = \{\underbrace{1,\ldots 1}_{k_1^{(i)}}, \ldots, \underbrace{2m, \ldots, 2m}_{k_{2m}^{(i)}} \}.
\end{split}
\end{equation}
In expression \eqref{eq:basis-change-Utheta}, $S_N$ represents the permutation group of degree $N$ and $O(\boldsymbol{\vartheta})$ is the (linear) mode basis change transformation at the level of $\mathcal A_{N = 1,m} = \{\hat q_1, \ldots, \hat q_m, \hat p_1, \ldots, \hat p_m\}$ (note that here there are only local operators). For a given number of modes $m$ the most general linear transformation $O$ that we are concerned with can be represented in a compact way, using, for example, Clements's decomposition \cite{Clements2017}, by the combination of $m(m-1)/2$ real beamsplitters and $m(m-1)/2$ local rotations (phase shifts), as represented in Fig.\ref{fig:Clements_decomposition}. The transformation depends on $m(m-1)$ parameters that we gather in the vector $\boldsymbol{\vartheta} = (\boldsymbol{\theta}, \boldsymbol{\varphi})$, where $\boldsymbol{\theta}$ contains the $m(m-1)/2$ angles of the beamspliters, and $\boldsymbol{\varphi}$ contains the $m(m-1)/2$ phases of the phase shifters. To reproduce the full set of passive transformations and additional set of $m$ local phase shifters should be included, nevertheless it is irrelevant for our task, given that these do not alter the entanglement property of the states we consider. More details on how to derive Eq. \eqref{eq:basis-change-Utheta} can be found in Appendix [REF].

To illustrate the quite complicated expression \eqref{eq:basis-change-Utheta},
let us take an example with $m=2$ and $N=3$: consider the generator $(2,0,1,0)$ [we decide to order the set of generators such that it corresponds to $i=3$] which corresponds to the symmetrised version of $\hat{q}_1^2 \, \hat{p}_1$. We then apply for simplicity the following transformation 
\begin{equation} \label{eq:O-Rtheta}
    O \equiv R(\theta) =  \begin{pmatrix}
        \cos \theta &  \sin \theta & 0 & 0 \\
         -\sin \theta & \cos \theta & 0  & 0 \\
          0 & 0 & \cos \theta &  \sin \theta  \\
      0 & 0 & -\sin \theta & \cos \theta  \\
    \end{pmatrix}.
\end{equation}
Let us derive the term $U_{31}(\theta)$, where $l=1$ corresponds to the generator $(1,1,1,0)$, i.e., the symmetrised version of $\hat{q}_1 \, \hat{p}_1 \, \hat{q}_2 $. In this case the vectors defined in Eq. \eqref{eq:alpha-beta-mode-basis-change} are $\alpha = (1,2,3)$ and $\beta = (1,1,3)$, such that it leads to 
\begin{equation*}
U_{31}(\theta) =   \frac{2}{1!}  O_{1 1}^{\rm \scriptscriptstyle T} \,  O_{2 1}^{\rm \scriptscriptstyle T} \, O_{3 3}^{\rm \scriptscriptstyle T} = 2 \cos^2 \theta  \, \sin \theta,
\end{equation*}
since all other terms are zero (it comes with a factor 2 since we need to sum over all permutations). By computing $(\cos \theta \hat{q}_1 + \sin \theta \hat q_2)^2 \, (\cos \theta \hat{p}_1 + \sin \theta \hat p_2)$, one can indeed check that the coefficient in front of $\hat{q}_1 \, \hat{p}_1 \, \hat{q}_2 $ is indeed given by $2 \cos^2 \theta  \, \sin \theta$. 

Expression \eqref{eq:basis-change-Utheta} is fully general for any transformation $O$ and any number of modes $m$ and any set of generators $\mathcal{A}_{\scriptscriptstyle N}$. However, the computation time increase rapidly when considering high order generators  given the factorial scaling with $N$.

\section{Application of the witness to experimentally relevant states}
\label{sec:application_witness}
To explore the capability of the witness derived in the previous section to detect mode intrinsic entanglement for bipartite and multimode systems, we use it on a specific class of non-Gaussian states: photon subtracted states. These states are produced by subtracting one or more photons from a Gaussian state.
The removal of photons is a non-Gaussian operation, thus rendering the state non-Gaussian. Therefore, it makes it possible to explore how quantum properties are affected by the non-Gaussianity using a class of states which falls within a well-established theoretical framework \cite{Walschaers2021, Walschaers2019} and has the advantage to be already produced in the lab with state-of-the-art experiments using different spatial and temporal modes of indistinguishable sources \cite{Neergaard-Nielsen2006, Wakui2007, Endo2023}, or in temporal-frequency domain \cite{Ra2020}. 

\begin{figure}[h!]
\includegraphics[width = \linewidth]{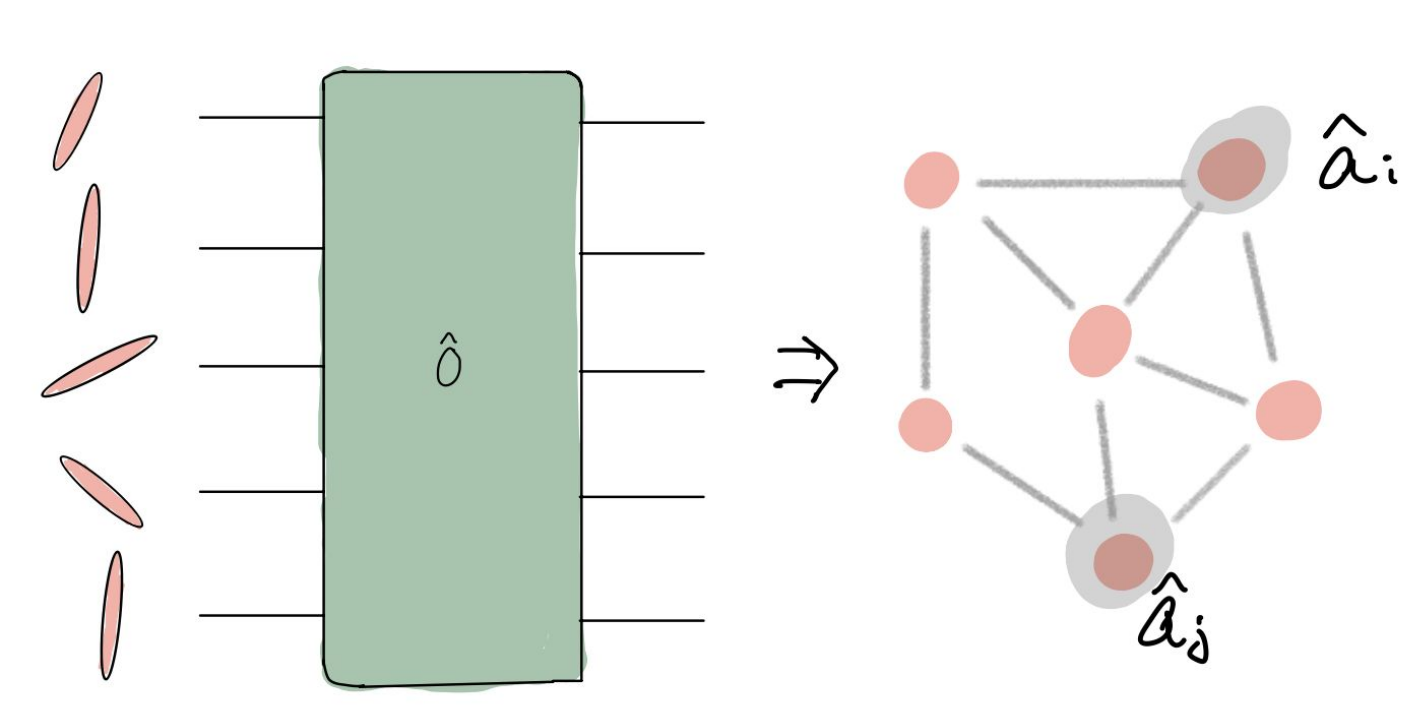}
    \caption{Graphical representation of the family of states that we consider. In the left we depict the cluster state as generated by applying an orthogonal transformations on a set of squeezed states. In the right we depict the cluster state by representing the links between the modes that interact directly along the transformation $\hat O$. The final state to build our probe states is to degaussify these clusters by performing photon subtraction in specific modes of our choice. In particular we will consider only up to two photon subtractions.}
    \label{fig:multimode_cluster}
\end{figure}
The general expression of the resulting states is given by
\begin{equation}\label{eq:pure_photon_subtracted_cluster}
    |\psi\rangle \propto \left(\prod_{i=1}^{N_S} \hat{a}_i \right)\hat{O} \left(\prod_{j=1}^{M} \hat{S}(r_j, \bar \phi_j)\right)|\mathbf 0\rangle,
\end{equation}
up to a normalization factor, where $N_S$ is the number of photons subtracted and $M$ is the number of modes in the cluster. $\hat{S}(r_j, \bar \phi_j)$ is a single mode squeezing operation that squeezes the quadrature $\hat{\xi}_{\bar \phi_j}=\hat{q}_j \cos \bar \phi_j +\hat p_j \sin \bar \phi_j$, of mode $j$, with a squeezing level $r_j$. $\hat O$ is a general passive transformation, and $\hat a_i$ are the annihilation operator corresponding to mode $i$. In Fig.~\ref{fig:multimode_cluster} we illustrate the case $N_s=2$. We also point out that photon subtractions can happen in a superposition of different modes of the cluster state; this case will only be considered for two-mode states, as in any case it can be casted into the form above by including the mixing operation in the unitary $\hat O$ (see Sec. \ref{sec:two-mode-photon-subtracted}).

We split this section in two subsections: we start with the case of multimode cluster states to show that the witness can be applied and successfully detect mode-intrinsic entanglement. In particular, we apply the metrological witness to three-mode, four-mode and five-mode cluster states.
Then we move to two-mode photon subtracted states. In some configurations discussed below this leads to mode-intrinsic entangled states that the witness can detect. For such a two-mode system the unitary transformation $U(\boldsymbol{\vartheta})$ to go from one basis to another (as represented in Fig.~\ref{fig:Clements_decomposition}) only depends on two parameters $\boldsymbol{\vartheta} = (\theta, \varphi)$. Therefore, one can easily visualize the metrological witness $E_{\rho}^{\mathcal A_{\scriptscriptstyle N,2}}(\theta, \varphi)$ defined in Eq. \eqref{eq:entanglement_witness_matrix_form} in all possible bases in a two-dimensional parameter space. 
Likewise, as we discuss in Sec. \ref{sec:relaxation_criteria}, the mode-intrinsic entanglement witness for two-mode photon subtracted states could be measured experimentally using the Fisher information matrix, making it a very interesting and promising study case towards the experimental detection of non-Gaussian entanglement. 

\subsection{Multimode CV-cluster states}

\begin{figure}[h!]
\centering
\includegraphics[scale = 1]{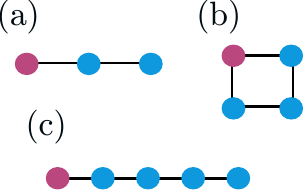}
    \caption{Photon subtracted (a) three-mode, (b) four-mode, (c) five-mode cluster states under consideration. Nodes represent modes and black edges represent entanglement links between modes. The red nodes indicate the mode where we subtract a photon. }
    \label{fig:cluster_states}
\end{figure}

A CV-cluster state is made of a set of modes which are coupled together by applying entangling $C_Z = \exp (i \,  \hat q_i \otimes \hat q_j)$ gates. Overall, they form a graph structure $(\mathcal{V}, \mathcal{E})$, where $\mathcal{V}$ is the set of nodes, i.e., the set of modes in the multimode state, while $\mathcal{E}$ represents the set of edges between the nodes, i.e., the quantum correlations that connect the modes together after applying the $C_Z$ gates. The graph formed by the cluster state can be described by the so-called adjacency matrix $V$ whose coefficients $V_{ij}$ are 1 if an edge exists between nodes $i$ and $j$ (in other words if modes $i$ and $j$ are entangled to each other), 0 otherwise. We show in Fig. \ref{fig:cluster_states} the three cluster states that we will consider in this section. As an example we give the adjacency matrix of the three-mode cluster state:
\begin{equation*}
    V_{\rm \scriptscriptstyle 3-modes} = \begin{pmatrix}
        0 & 1 & 0 \\
        1 & 0 & 1 \\
        0 & 1 & 0
    \end{pmatrix}.
\end{equation*}
The red nodes in Fig. \ref{fig:cluster_states} indicate in which mode we subtract a photon to create our photon subtracted cluster state. 

A cluster state associated with a given adjacency matrix $V$ is obtained from a set of single-mode squeezed states by applying a unitary matrix 
\begin{equation} \label{eq:unitary-cluster}
    U_{\scriptscriptstyle V} = (\mathds{1} + i \, V)\, (V^2 + \mathds{1})^{-1/2} \, \mathcal{O},
\end{equation}
where $\mathds{1}$ is the unitary matrix and $\mathcal{O}$ can be any real orthogonal matrix \cite{vanloock2007, ferrini2015, sansavini2020}. The choice of $\mathcal{O}$ can be optimized to target a specific cluster configuration as discussed in Ref. \cite{ferrini2015}. Indeed, a perfect cluster state where correlations between modes are exactly given by the adjacency matrix can only be obtained in the limit of infinite squeezing. In this case, the variances of certain combinations of quadratures, called nullifiers, are exactly zero; these nullifiers are defined as
\begin{equation}
    \hat \delta_i \equiv \hat p_i  - \sum_{k=1}^m V_{ik} \hat q_k, \quad \forall i \in \{1, \ldots, m \},
\end{equation}
where $m$ is the number of modes in the cluster. In the case of finite squeezing, the variances of the nullifiers are not zero but one can minimize them to tend to the limit of the perfect cluster state by optimizing the orthogonal matrix $\mathcal{O}$ in Eq. \eqref{eq:unitary-cluster}. We thus follow the procedure described in Ref. \cite{ferrini2015} using a genetic algorithm to minimize the variances of the nullifiers and obtain the best cluster states given the three graph structures of Fig.~\ref{fig:cluster_states}.
In order to consider realistic squeezing parameters we take the ones measured in the 4-mode cluster states of Ref. \cite{Medeiros2014}:  $- 7$ dB, $- 6$ dB, $-4$ dB, 0 dB. For the other cluster states considered here we use similar squeezing parameters:  $- 7$ dB, $-4$ dB, 0 dB (3-mode cluster state); $- 7$ dB, $- 6$ dB, $-4$ dB, $-2$ dB, 0 dB (5-mode cluster state). 

Before we subtract a photon, the cluster states are fully Gaussian. Therefore, we do not expect to witness any mode-intrinsic entanglement since there will always exist a basis where the states are separable. The subtraction of a photon in the first mode (the red nodes in Fig.~\ref{fig:cluster_states}) renders the state non-Gaussian and will affect the modes in its vicinity; we thus anticipate to detect mode-intrinsic entanglement.
In the case of a $m$-multimode system we have the freedom to choose how we split it into $p$ different partitions $\mathcal{P}_1|\ldots|\mathcal{P}_p$ and eventually detect $p$-partite mode-intrinsic entanglement. It is important to note that, in our case, the labelling of the modes is irrelevant; only the number of modes in each partition matters. Indeed, for any given mode splitting, we can always find a linear passive operation that changes the partitioning while keeping the same number of modes in each partition $\mathcal{P}_i$ (but with different mode labels). Therefore, as the witness covers all possible mode bases, any two mode splittings differing only by a reordering of the mode labels are completely equivalent regarding mode-intrinsic entanglement.

Tables \ref{tab:cluster3}, \ref{tab:cluster4}, \ref{tab:cluster5} gather the values of the mode-intrinsic entanglement witness $\mathcal{W}_Q(\hat \rho, \mathcal{A}_{N=1,m})$ -- using first-order generators only -- obtained for the three clusters and for all possible mode splittings (up to a reordering of the mode labels as discussed above). Given the number of parameters that we need to optimize to find the minimum of the entanglement witness $E_{\rho}^{\mathcal A_{\scriptscriptstyle N}}(\boldsymbol{\vartheta})$ (see Fig. \ref{fig:Clements_decomposition} where each blue and red block of the unitary transformation adds one degree of freedom), we also use here a genetic algorithm to solve this optimization problem. As we can clearly see from the results, for many mode splittings, the subtraction of the photon renders the cluster states mode-intrinsic entangled. On the other hand, it is very instructive to notice that only in the case where we split the cluster states into a bipartition where one of the subsystems contain $m-1$ modes, the value of the witness becomes zero. In this case, while we cannot guarantee that the cluster is not entangled in at least one basis since we only use a witness of entanglement, the non-Gaussian correlations can be probably localized in the subsytem containing $m-1$ modes leaving only Gaussian correlations between both subsystems.

\renewcommand{\arraystretch}{1.5}
\begin{table}[h!]
    \centering
    \includegraphics[scale = 0.7]{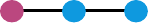} \\
    \begin{tabular}{|c|c|c|}
    \hline
        Partitions & $1|2|3$ & $12|3$  \\
        \hline
       $\mathcal{W}_Q(\hat \rho, \mathcal{A}_{N=1,3})$ & 0.94 & 0
         \\ \hline
    \end{tabular}
    \caption{Values of the mode-intrinsic entanglement witness derived in Eq. \eqref{eq:PS-metrological-witness} for the 3-mode cluster state considered in Fig. \ref{fig:cluster_states}~(a) and for all possible partitions (up to permutations).}
    \label{tab:cluster3}
\end{table}

\begin{table}[h!]
    \centering
    \includegraphics[scale = 0.7]{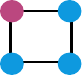} \\
    \begin{tabular}{|c|c|c|c|c|}
    \hline
        Partitions & $1|2|3|4$ & $12|3|4$ & $12|34$ & $1|234$  \\
        \hline
       $\mathcal{W}_Q(\hat \rho, \mathcal{A}_{N=1,4})$ & 0.87 & 0.33 & 0.33 & 0
         \\ \hline
    \end{tabular}
    \caption{Values of the mode-intrinsic entanglement witness derived in Eq. \eqref{eq:PS-metrological-witness} for the 4-mode cluster state considered in Fig. \ref{fig:cluster_states}~(b) and for all possible partitions (up to permutations).}
    \label{tab:cluster4}
\end{table}

\begin{table}[h!]
    \centering
    \includegraphics[scale = 0.7]{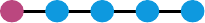} \\
    \begin{tabular}{|c|c|c|c|c|c|c|}
    \hline
        Partitions & $1|2|3|4|5$ & $12|3|4|5$ & $12|34|5$  & $123|4|5$  & $123|45$ & $1234|5$  \\
        \hline
       $\mathcal{W}_Q(\hat \rho, \mathcal{A}_{N=1,5})$ & 0.92 & 0.31 & 0.31 & 0.17 & 0.17 & 0
         \\ \hline
    \end{tabular}
    \caption{Values of the mode-intrinsic entanglement witness derived in Eq. \eqref{eq:PS-metrological-witness} for the 5-mode cluster state considered in Fig. \ref{fig:cluster_states}~(c) and for all possible partitions (up to permutations).}
    \label{tab:cluster5}
\end{table}

While it is difficult to draw any systematic conclusion from the results above given that we only tested the witness in three specific cluster configurations, it shows that the mode-intrinsic entanglement witness with first-order generators is able to detect mode-intrinsic entanglement for experimentally relevant cluster states.
It also opens a new toolbox to explore and better understand the entanglement structure of large multimode non-Gaussian states such as those considered in Ref. \cite{Walschaers2023Network}. A deeper exploration of this tool for the study of large clusters is left for future works. In the rest of this paper we rather focus on the study of the smallest version of these systems, \ie two mode, two photon subtracted states. In this more controlled setting we can fully explore the capabilities and limitations of the method we develop, while also restricting to a scenario with more immediate potential to be tested experimentally. 


\subsection{Two-mode photon subtracted states}\label{sec:two-mode-photon-subtracted}

As explained above, we now turn our attention to two-mode photon subtracted states defined by 
\begin{equation}\label{eq:two-photon-subt-states}
    |\psi \rangle = \prod_{i=1}^{N_{\scriptscriptstyle S}} (\cos \Theta_i \,  \hat a_1 + \sin \Theta_i \,  \hat a_2)  \, \hat S(r_1) \, \hat S(r_2) \, |00\rangle,
\end{equation}

where $\hat S(r_1)$ and $\hat S(r_2)$ are single-mode squeezing operators, equivalent to those in equation \eqref{eq:pure_photon_subtracted_cluster}, where we removed the phase dependence $\bar \phi_i$ and allow $r_i$ to be negative. Angles $\Theta_i$ control mode superposition in which photons are subtracted, which effectively modifies the probability that the subtracted photon came from either mode for each subtraction $j = 1, \ldots, N_{\scriptscriptstyle S}$. In general we could have considered the subtraction to happen in an already entangled Gaussian state, nevertheless, those can always be brought to the form in equation \eqref{eq:two-photon-subt-states} by the action of some passive unitary. In this subsection we consider two cases: $N_{\scriptscriptstyle S}= 1$ and $N_{\scriptscriptstyle S}= 2$. In both cases we consider the single mode squeezed states to be either squeezed in position ($r_i >0$) or in momentum ($r_i<0$). 
One-photon subtracted states ($N_{\scriptscriptstyle S}= 1$) are not mode-intrinsic entangled in the two following cases: (i) $r_1 = r_2$ and $\forall \, \Theta_1 $, (ii) $\Theta_1 = n \pi/2$, $n \in \mathbb{Z}$ and $\forall r_1, r_2$; and, similarly, two-photon subtracted ($N_{\scriptscriptstyle S}= 2$) are not mode-intrinsic entangled for: (i) $r_1 = r_2$ and $\forall \, \Theta_1 $, (ii) $\Theta_1 = n \pi/2$, $n \in \mathbb{Z}$ and $\forall r_1, r_2$ ($N_{\scriptscriptstyle S}= 1$); (i) $r_1 = r_2$, $\Theta_1$ and $\Theta_2$ orthogonal to each other, (ii) $\Theta_1 = n \pi/2$ and $\Theta_2 = m \pi/2$ $n,m \in \mathbb{Z}$ and $\forall r_1, r_2$.
In all other cases one can prove that these states are entangled in all mode bases. To illustrate how the metrological witness \eqref{eq:PS-metrological-witness} can detect mode intrinsic entanglement we now consider two specific examples:
 $N_{\scriptscriptstyle S} = 1$, $r_1 = -r_2 = 0.2$, $\Theta_1 = \pi/4$; $N_{\scriptscriptstyle S} = 2$, $r_1 = -r_2 = 0.2$, $\Theta_1 =- \Theta_2 = \pi/4$.

In the case of a two-mode state, the mode basis change transformation at the level of quadrature operators using Clement's decomposition can be written as 
\begin{equation}
    O(\theta, \varphi) = R(\theta) \,  Ph(\varphi), 
\end{equation}
where $R(\theta)$ is given in Eq. \eqref{eq:O-Rtheta} and
\begin{equation*}
   Ph(\varphi) =
   \begin{pmatrix}
          \cos \varphi & 0 &  \sin \varphi  & 0 \\
           0 & 1 & 0  & 0 \\
      -\sin \varphi & 0 & \cos \varphi & 0  \\
      0 & 0 & 0 & 1  \\
         \end{pmatrix}.
\end{equation*}
We can then compute the mode basis transformation $U(\theta, \varphi)$ using Eq. \eqref{eq:basis-change-Utheta} and apply the metrological witness $E_{\rho}^{\mathcal A_{\scriptscriptstyle N}}(\theta, \varphi)$ [see Eq. \eqref{eq:PS-metrological-witness}] using first order and second order generators $\mathcal{A}^{\scriptscriptstyle \rm (L)}_{N=1}$ and $\mathcal{A}^{\scriptscriptstyle \rm (L)}_{N=2}$, respectively. Notice that we have omitted the subscript $m=2$ from the definition of the set of generators, as from this point on we will only be dealing with two-mode states. 

\begin{figure}[h!]
\includegraphics[width = \linewidth]{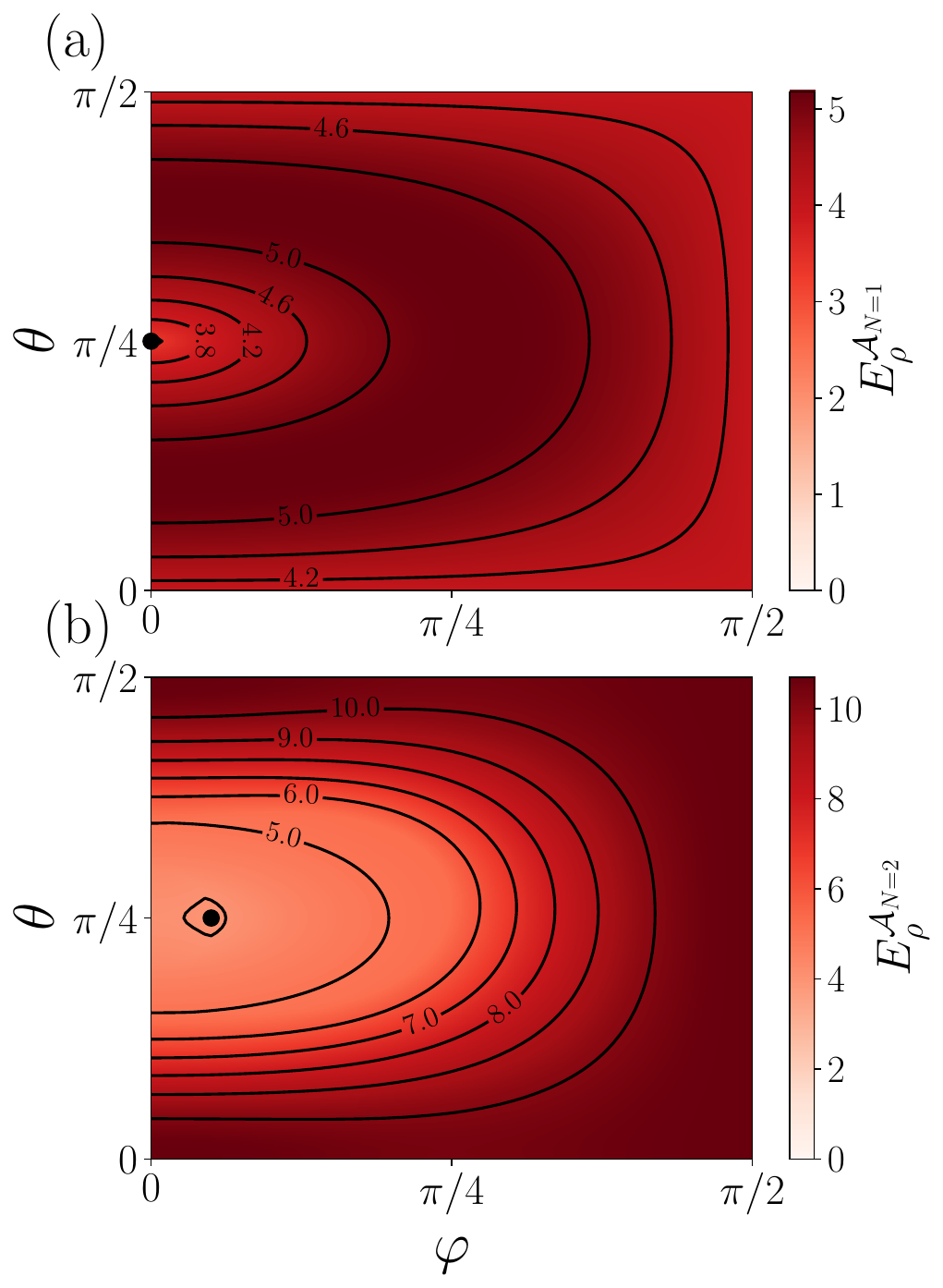}
    \caption{Entanglement witness (a) $E_{\rho}^{\mathcal A_{\scriptscriptstyle N = 1}}(\theta, \varphi)$ for the first order set of generators $\mathcal A_{\scriptscriptstyle N = 1}$ and (b)  $E_{\rho}^{\mathcal A_{\scriptscriptstyle N = 2}}(\theta, \varphi)$ for generators up to second order ($\mathcal A_{\scriptscriptstyle N = 2}$) as a function of the mode basis change parameters $(\theta, \varphi)$. Here we consider a one photon subtracted state with $r_1 = -r_2 = 0.2$, $\Theta_1 = \pi/4$. The black dots indicate the minimal value over the parameter space.}
    \label{fig:1phSub}
\end{figure}

\begin{figure}[h!]
\includegraphics[width = \linewidth]{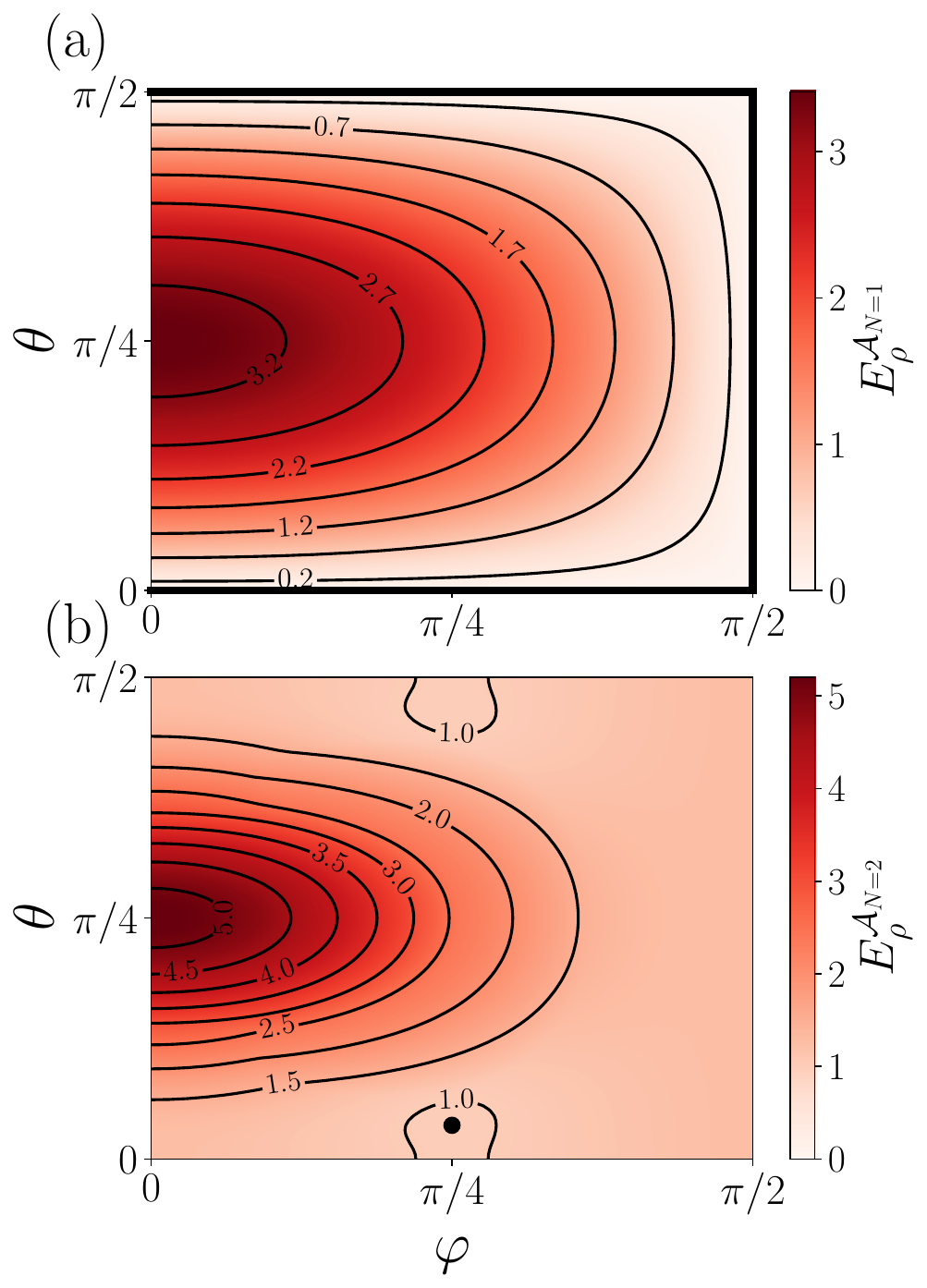}
    \caption{Entanglement witness (a) $E_{\rho}^{\mathcal A_{\scriptscriptstyle N = 1}}(\theta, \varphi)$ for the first order set of generators $\mathcal A_{\scriptscriptstyle N = 1}$ and (b)  $E_{\rho}^{\mathcal A_{\scriptscriptstyle N = 2}}(\theta, \varphi)$ for generators up to second order ($\mathcal A_{\scriptscriptstyle N = 2}$) as a function of the mode basis change parameters $(\theta, \varphi)$. Here we consider a two photon subtracted state with $r_1 = -r_2 = 0.2$, $\Theta_1 =- \Theta_2 = \pi/4$. The black lines for (a) and the black dot for (b) indicate the minimal value over the parameter space.}
    \label{fig:2phSub}
\end{figure}

\begin{figure}[h!]
\centering
\includegraphics[width =\linewidth]{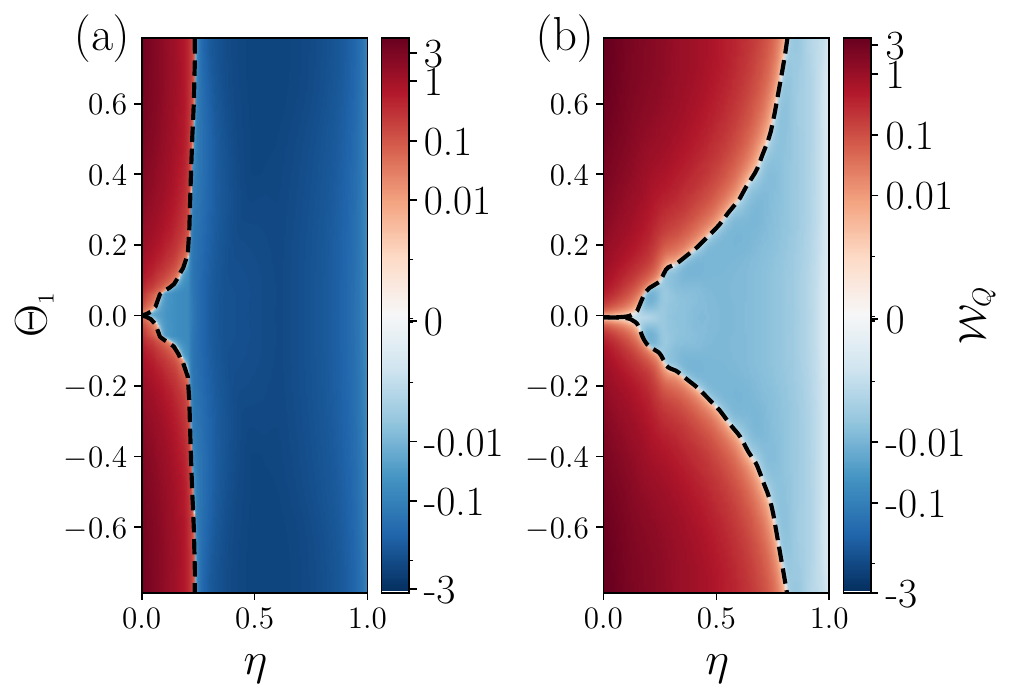}
    \caption{Metrological witness $\mathcal{W}_Q$ [see Eq. \eqref{eq:PS-metrological-witness}] for a single photon subtracted state ($N_{\scriptscriptstyle S} = 1$) with $r_1 = -r_2 = 0.2$. The witness is computed (a) using the set of first order generators $\mathcal{A}_{\scriptscriptstyle N = 1}$ and (b) using generators up to second order ($\mathcal{A}_{\scriptscriptstyle N = 2}$), as a function of the optical efficiency $\eta$ and the angle of photon subtraction $\Theta_1$ ranging from $-\pi/4$ to $\pi/4$. The black dashed lines delimit the region of negative values (blue area) from the region of positive values (red area). The red area corresponds to the region where the state is detected to be not passively separable.}
    \label{fig:lossy_1photonSubState_BothOrder}
\end{figure}

\begin{figure}[h!]
\centering
\includegraphics[width =\linewidth]{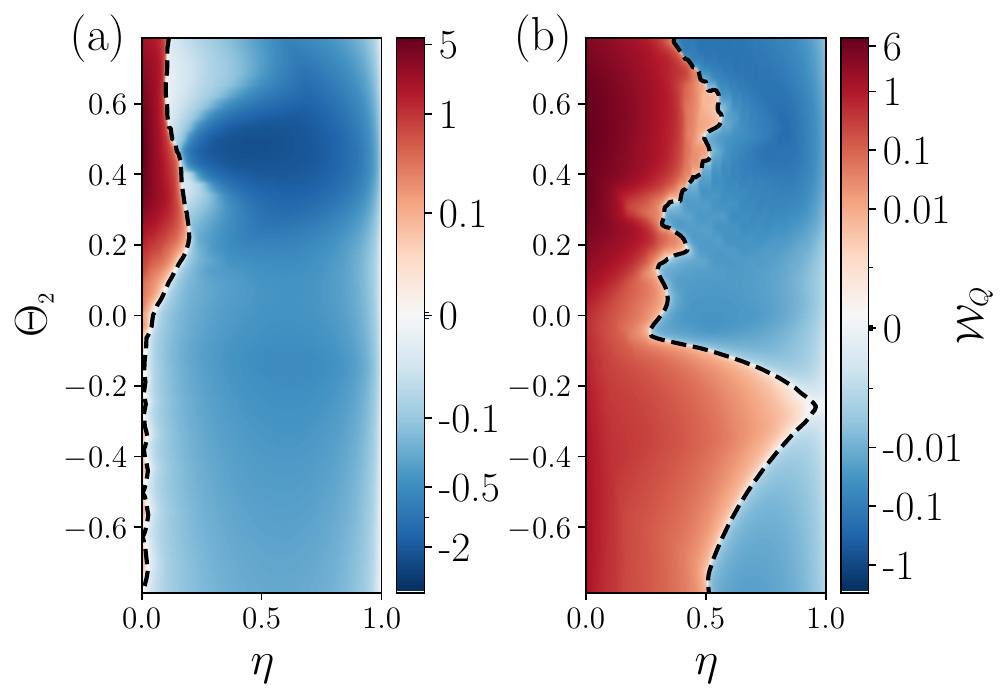}
    \caption{Metrological witness $\mathcal{W}_Q$ [see Eq. \eqref{eq:PS-metrological-witness}] for a two photon subtracted state ($N_{\scriptscriptstyle S} = 2$) with $r_1 = 1.5$ dB, $r_2 = -2.6$ dB and $\Theta_1 = \pi/4$. The witness is computed (a) using the set of first order generators $\mathcal{A}_{\scriptscriptstyle N = 1}$ and (b) using generators up to second order ($\mathcal{A}_{\scriptscriptstyle N = 2}$), as a function of the optical efficiency $\eta$ and the angle of the second photon subtraction $\Theta_2$ ranging from $-\pi/4$ to $\pi/4$. The black dashed lines delimit the region of negative values (blue area) from the region of positive values (red area). The red area corresponds to the region where the state is detected to be not passively separable. }
    \label{fig:lossy_2photonSubState_BothOrder_r1_1Pt5dB}
\end{figure}

The results are shown  in Fig. \ref{fig:1phSub} for $N_{\scriptscriptstyle S} = 1$ and Fig. \ref{fig:2phSub} for $N_{\scriptscriptstyle S} = 2$. In both figures, the top graphs labeled as (a) correspond to the results using generators in the set $\mathcal{A}_{N=1}$, while we used the set $\mathcal{A}_{N=2}$ to produce the bottom graphs labeled as (b). The black dots indicate the minimal value of $E_{\rho}^{\mathcal A_{\scriptscriptstyle N}}(\theta, \varphi)$ when spanning the parameter space $(\theta, \varphi)$. One sees that for the one-photon subtracted state the metrological witness is able to detect mode intrinsic entanglement both at first order and second order on the generators: the minimal value over the parameter space is always strictly greater than zero, thus demonstrating entanglement in all mode bases. The robustness of the witness against optical losses is adressed below (see in particular Figs. \ref{fig:lossy_1photonSubState_BothOrder} and \ref{fig:lossy_2photonSubState_BothOrder_r1_1Pt5dB}). 
In the case of the two-photon subtracted state considered here the situation is different; as shown in Fig. \ref{fig:2phSub} (a), first order fails to detect entanglement in all bases: the minimum of $E_{\rho}^{\mathcal A_{\scriptscriptstyle N = 1}}(\theta, \varphi)$ reaches zero for a range of values $\theta = 0$ $(\forall \phi)$, $\theta = \pi/2$ $(\forall \phi)$ and $\phi = \pi/2$ $(\forall \theta)$ (see the black lines at the borders of the figure).
Therefore, the first-order protocol cannot be used to assess the mode intrinsic entanglement of the state. On the contrary, as shown in Fig. \ref{fig:2phSub} (b), once we use second order generators, the minimal value when spanning all mode-bases become strictly positive (it equals 0.98 in this precise example), thus meaning that the state is mode-intrinsic entangled. This illustrates the interest to use higher generators and to not restrict the metrological witness to first order generators.
More generally one can show that for two photon subtracted states ($N_{\scriptscriptstyle S} = 2$) first order always fails when $\Theta_1$ and $\Theta_2$ are orthogonal to each other.  

So far we did not include optical losses inherent to any experimental setup. We can thus reproduce the previous results and apply the metrological witness to lossy one and two-photon subtracted states. To simulate such losses we apply two single-mode loss channels $\hat{L}_1(\eta)$ and $\hat{L}_2(\eta)$ to each mode as described in Ref. \cite{Eaton2022}, where $\eta$ is the efficiency coefficient: $\eta = 0$ means 100\% loss, while $\eta = 1$ means no loss. 
Figs. \ref{fig:lossy_1photonSubState_BothOrder} (a) and (b) show the value of $\mathcal W_{\mathcal Q}(\hat \rho,\mathcal A_{N})=\underset{\varphi, \alpha}{\text{min}} \, E_{\rho}^{\mathcal A_{\scriptscriptstyle N}}(\theta, \varphi)$ for first and second order generators ($N=1$ and $N=2$) as a function of the optical efficiency $\eta$ and of the angle $\Theta_1$ for a one photon subtracted state with $r_1 = -r_2 = 0.2$. We recall that $\Theta_1$ parameterizes the basis where we subtract the photon.
Figs. \ref{fig:lossy_2photonSubState_BothOrder_r1_1Pt5dB} (a) and (b) also show the value of $\mathcal W_{\mathcal Q}(\hat \rho,\mathcal A_{N})$ for $N=1$ and $N=2$ as a function of the optical efficiency $\eta$ and of the angle of the second photon subtraction $\Theta_2$ for a two photon subtracted state with $r_1 = 1.5$ dB, $r_2 = -2.6$ dB and $\Theta_1=\pi/4$. This choice of squeezing parameters is motivated from the results obtained in Ref. \cite{David_2023} where it has been observed that such squeezing parameters lead to a better resilience to losses. 
The black dashed curve delimits the regions of positive (red area) and negative (blue area) values. The red area thus corresponds to the region where the state is detected to be not passively separable. 
As one can see from the figures, the metrological witness is able to detect mode-intrinsic entanglement even in the presence of losses. However, one sees immediately that using second order generators results in a much better resilience to losses compared to first order both for one and two photon subtracted states; it thrives to detect entanglement in all mode bases for a wide range of losses (up to 95 $\%$ of losses in the case of two-photon subtracted state when $\Theta_2 \simeq -0.25$). 

Of course the maximal amount of losses to detect intrinsic mode entanglement shown in Figs. \ref{fig:lossy_1photonSubState_BothOrder} and \ref{fig:lossy_2photonSubState_BothOrder_r1_1Pt5dB} (black dashed lines) is a theoretical bound and represents more a theoretical case study to demonstrate the robustness of the witness against losses: indeed, we used here the Quantum Fisher information matrix $\mathcal{Q}_{\hat \rho}^{\mathcal A_{\scriptscriptstyle N}}$ to evaluate the metrological criterion \eqref{eq:PS-metrological-witness}. Such a quantity is difficult if not impossible to measure experimentally; what can be measured though is the Fisher information matrix (bounded by the Quantum Fisher information), resulting in a weaker version of the mode-intrinsic-entanglement metrological-witness. However, even in this case, we show in Sec. \ref{sec:relaxation_criteria} that the witness is able to detect entanglement in all mode bases while considering quantities that are experimentally accessible. 

\section{Relaxation of the witness}\label{sec:relaxation_criteria}

So far we have considered a protocol that relies on the Quantum Fisher information matrix. This is a strong requirement if we want to design a protocol that aims to be applicable in experimental settings since it requires full state tomography, an unfeasible task in a multimode setting. Nevertheless for a given measurement setting we know that the Fisher information is a lower bound of the quantum Fisher information [see Eq. \eqref{eq:comp_FI_QFI}]. Therefore, one can derive a lower bound -- but experimentally accessible -- of the mode-intrinsic entanglement witness introduced in the previous sections by relying on the (classical) Fisher information matrix -- instead of the Quantum Fisher information matrix --  whose elements will be defined below in Eq. \eqref{eq:class-FIM-elements}. \par
\begin{figure}
\includegraphics[width =\linewidth]{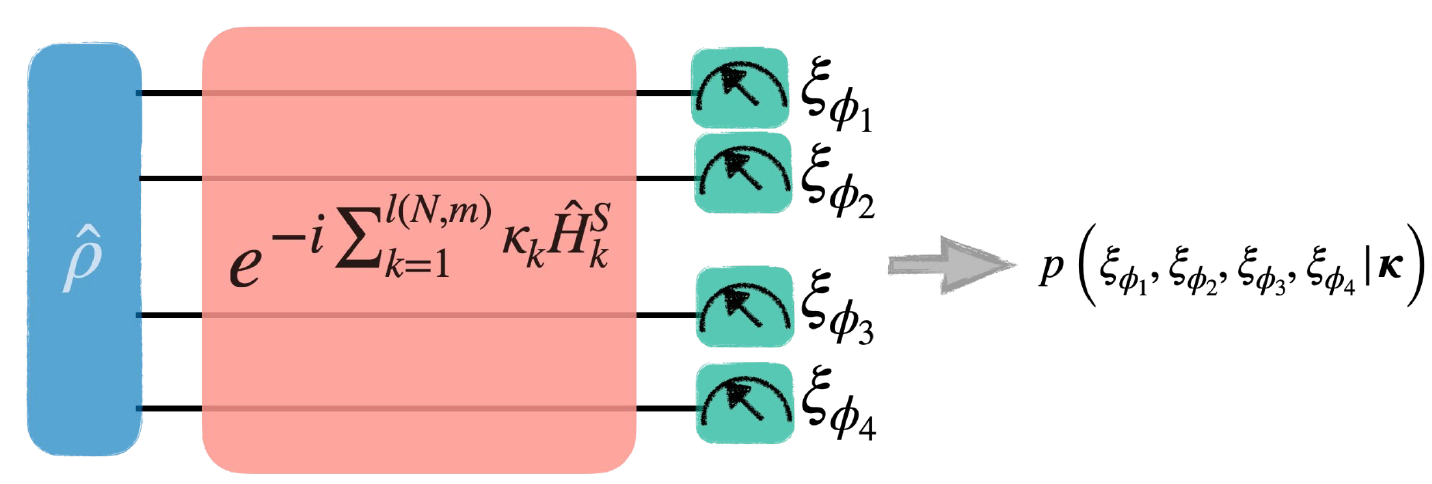}
    \caption{Measurement setting considered for the estimation of a lower bound on the quantum Fisher information matrix. The joint measurement of quadratures in the different modes allows to build a probability distribution of outcomes $p(\xi_{\phi_1},...,\xi_{\phi_N})$. From the influence of the different generators in this object we can estimate the associated classical Fisher information matrix. }
    \label{fig:homodyne-detection}
\end{figure}

To proceed, as illustrated in Fig.~\ref{fig:homodyne-detection}, we consider a setting in which we implement, either physically or, if possible, in post-processing, a transformation $\exp{(-i \sum_{k=1}^{l(N, m)} \kappa_k \hat H_k^{\mathcal{S}})}$ on a quantum state $\hat \rho$. We recall that the generators $\hat H_k^{\mathcal{S}}$ are the symmetrised operators defined in Eq. \eqref{eq:generators-quadratures} and $l(N, m)$ is the length of the set containing all generators of order $N$ for a $m$-mode system [see expression \eqref{eq:length_set}]. Each generator implements a parameter $\kappa_j$ that we gather in a vector $\boldsymbol{\kappa} = (\kappa_1,...,\kappa_{\ell(N,m)})$.
Then, we perform homodyne detection in each mode $i$ of the system in quadratures $\hat \xi_{\phi_i} = \cos \phi_i \hat q_i + \sin \phi_i \hat p_i$ with $\phi_i \in\left[0,\pi\right)$, i.e., we consider measurement projectors $\hat \Pi_i = |\xi_{\phi_i}\rangle \langle \xi_{\phi_i}|$, where $|\xi_{\phi_i}\rangle$ is an eigenstate of the quadrature operator $\hat \xi_{\phi_i}$.
The elements of the classical Fisher information matrix for a given state $\hat \rho$ and the measurement setting described above are computed as 
\begin{equation}\label{eq:class-FIM-elements}
    \begin{split}
&F(\hat{\rho},\boldsymbol{\kappa})_{i,j}=\int_{\Re^{m}}\left(\Pi_{i=1}^{m}d \xi_{\phi_i}\right)p(\xi_{\phi_1}, \ldots, \xi_{\phi_m}|\boldsymbol{\kappa}=\mathbf{0})\\
    &\qquad \qquad \qquad\frac{\partial \log p(\xi_{\phi_1}, \ldots, \xi_{\phi_m}|\kappa_1,\ldots,\kappa_{l((N,m)})}{\partial \kappa_i }|_{\boldsymbol{\kappa}=\mathbf{0}}\\
    &\qquad \qquad \qquad\frac{\partial \log p(\xi_{\phi_1}, \ldots, \xi_{\phi_m}|\kappa_1, \ldots,\kappa_{l((N,m)})}{ \partial \kappa_j}|_{\boldsymbol{\kappa}=\mathbf{0}}
    \end{split}
\end{equation}
where $p(\xi_{\phi_1}, \ldots, \xi_{\phi_m}|\kappa_1, \ldots ,\kappa_{l((N,m)})$ corresponds to the probability distribution of homodyne outcomes after all the parameters $\mathbf{\kappa}$ are implemented by the corresponding generators. To compute this quantity we need to know how each generator in our set acts on the probability distribution. Both theoretically and experimentally the problem is feasible only up to generators of order two. For two modes this set is given by $\mathcal{A}^{\rm \scriptscriptstyle (NL)}_{\scriptscriptstyle N = 2} $ in Eq. \eqref{eq:set-generators-order-2}. All of them correspond to symplectic transformations or displacements that can be implemented theoretically by a change of coordinates in the Wigner function, 
\begin{equation}\label{eq:Wigner_funct_transform_symp_gen}
W(\vec{\xi} \, |\kappa_j)=W(S_j^{-1}  \, \vec{\xi}-\vec{\alpha}_j),
\end{equation}
where $S_j$ and $\vec \alpha_j$ correspond to the symplectic transformation and displacements, respectively, associated to the generator $\hat H_j^{\mathcal{S}}$ contained in $\mathcal{A}^{NL}_{N=2}$, while $\vec \xi = (q_1, \ldots, q_m, p_1, \ldots, p_m)$ gather all quadratures.
Once the Wigner function is transformed, to obtain the probability distribution of outcomes for the measurement setting that we consider, we can, at the level of the Wigner function rotate each quadrature by $\phi_i, \, i \in \{1, \ldots, m\}$, i.e., 
\begin{equation}
    W(\vec{\xi}_{\phi} \, |\kappa_j) = W\left[ \mathcal{R}(\phi) \left(  S_j^{-1}  \,\vec{\xi}-\vec{\alpha}_j \right)\right],
\end{equation}
where $\mathcal{R}(\phi)$ is a rotation operator that transforms the set of quadratures $(q_1, \ldots, q_m,\,  p_1, \ldots, p_m)$ to $\vec{\xi}_{\phi} \equiv (\xi_{\phi_1}, \ldots ,\xi_{\phi_m},\, \overline{\xi}_{\phi_1} , \ldots,  \overline{\xi}_{\phi_m})$, with $\overline{\xi}_{\phi_i} = -\sin \phi_i q_i + \cos \phi_i p_i$.
As a final step, we just need to marginalize to the $m$ quadratures of interest to obtain the corresponding parametrized probability distribution
\begin{equation}\label{eq:marginal_prob_distrib_parametrized}
    p(\xi_{\phi_1}, \ldots,\xi_{\phi_m}|\kappa_j)=\int_{\Re^{m}}\Pi_{i=1}^{m}d \overline{\xi}_{\phi_i} W(\vec{\xi}_{\phi}|\kappa_j).
\end{equation}

From an experimental point of view symplectic operations are expected to be feasible \cite{Sun2023,Assad_2020,Andersen_2024}, even though currently challenging to implement.  On the other hand they are a necessary requirement for universal continuous variable quantum computation \cite{Bourassa2021, Andersen_2024}. Even though we are not currently capable to experimentally implement all of these unitaries, we should keep in mind that some of them can also be implemented in postprocessing of the measurement data \cite{Ferrini_2013}. The most simple example is the case of displacements, generated by the quadrature operators (\ie at order one), for example 
\begin{equation}\label{eq:effect_displacements}
p(\xi_{\phi_1}, \ldots,\xi_{\phi_m}|\kappa_{m+1})=p(\xi_{\phi_1}-\kappa_{m+1} \cos(\phi_1), \ldots ,\xi_{\phi_m}),
\end{equation}
where $\kappa_{m+1}$ is the parameter generated by the $(m+1)$-th generator contained in $\mathcal{A}_{N=1}^{NL}$, i.e., $\hat{p}_{1}$. At order 2 the only generators that cannot be treated in postprocessing are the shearing, \ie those of the form $\hat q_{i}^{2}$, whose effect is given by 
\begin{equation}
W(\vec{\xi}|\kappa_{\hat q_{i}^2})=W(p_{1}, \ldots ,q_{m},p_1, \ldots,p_{i}-2\kappa_{\hat q_{i}^2} q_{i}, \ldots ,p_{m}),
\end{equation}
which mixes quadratures that cannot be simultaneously measured. \par 

\begin{figure*}
\includegraphics[width =\textwidth]{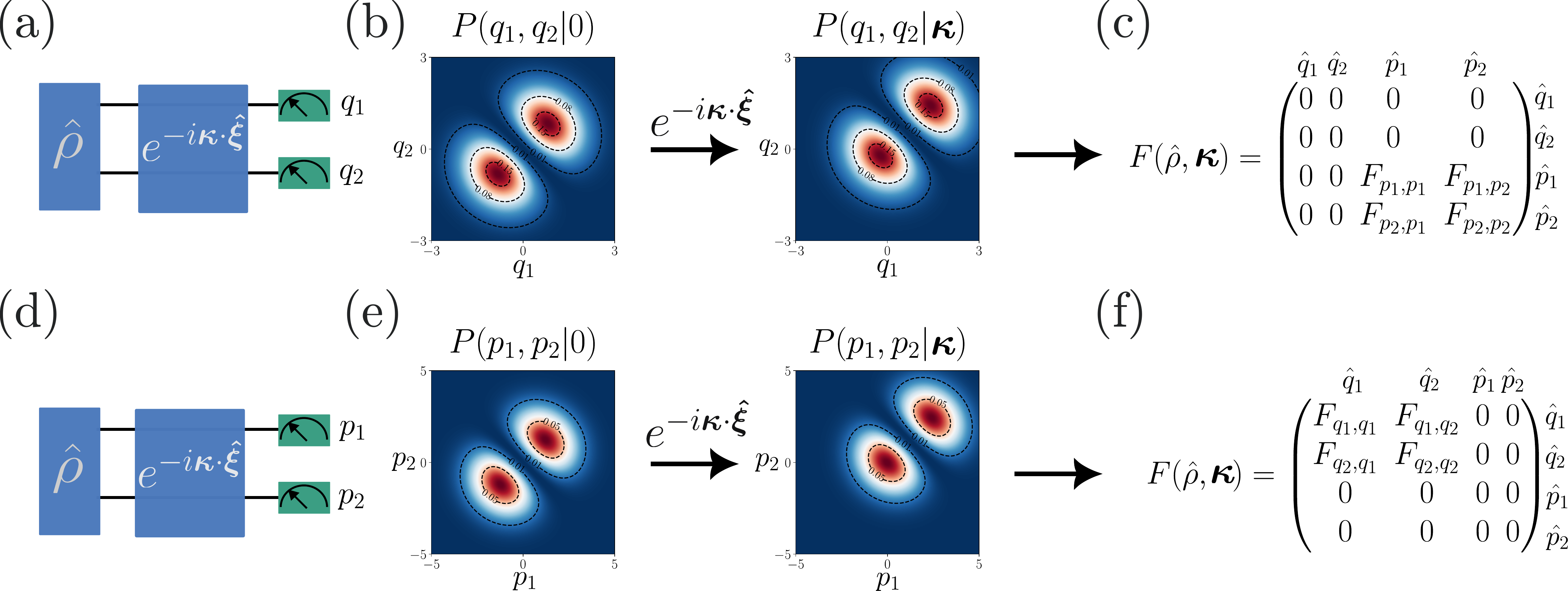}
    \caption{Schematic procedure to obtain the Fisher information matrix for different measurement settings in the case of a one photon subtracted state and first order generators. (a) Schematic representation of the metrological parameter estimation procedure generated by operators contained in the vector $\boldsymbol{\hat \xi} = (\hat q_1, \hat q_2, \hat p_1, \hat p_2)$, i.e., $\mathcal{A}_{N=1}$, and whose parameters are gathered in $\boldsymbol{\kappa} = (\kappa_1, \kappa_2, \kappa_3, \kappa_4)$. The measurement setting consists of homodyne detection along the position quadratures $q_1$ and $q_2$. (b) Effect of the displacement transformation $\exp{(-i \,  \boldsymbol{\kappa}  \cdot \boldsymbol{\hat \xi})}$ on the probability distribution $p(q_1, q_2)$. (c) Resulting Fisher information matrix. (d) Same schematic representation of the metrological protocol in the measurement basis $(p_1, p_2)$. (e) Effect of the displacement transformation $\exp{-i \sum_{k=1}^{4} \kappa_k \, \hat \xi_k}$ on the probability distribution $p(p_1, p_2)$. (f) Resulting Fisher information matrix.  }
    \label{fig:Fisher_setting}
\end{figure*}

We can then construct the Fisher information matrix $F(\hat{\rho},\boldsymbol{\kappa})$ by inserting the resulting probability distributions \eqref{eq:marginal_prob_distrib_parametrized} in a given measurement setting $\vec \xi_{\phi}$ in Eq. \eqref{eq:class-FIM-elements}.
Each choice of measurements gives rise to a different Fisher information matrix. To illustrate this point we refer to Fig. \ref{fig:Fisher_setting} which shows how this matrix is affected by two different choices of measurement settings for a one-photon subtracted state. For the sake of simplicity, we restrict ourselves to two modes and to displacement operators, i.e.,  first order generators contained in the set $\mathcal{A}_{N=1} = (\hat q_1, \hat q_2, \hat p_1, \hat p_2)$ (denoted as the vector operator $\hat \xi$ in Fig. \ref{fig:Fisher_setting}). In the first scenario we measure position quadratures $q_1$ and $q_2$ (see the left top scheme of Fig. \ref{fig:Fisher_setting}), which results in the probability distributions $p(q_1, q_2|0)$ and $p(q_1, q_2|\boldsymbol{\kappa})$ (shown in the middle top plots) before and after we apply the transformation on the state. Note that for displacement operators the transformation can be applied in post-processing, i.e., directly from the probability distribution $p(q_1, q_2|0)$ reconstructed from experimental data. When measuring the quadratures $(q_1, q_2)$ the probability distribution of outcomes is only affected by displacement operators $\hat p_1$ and $\hat p_2$, resulting in non-zero elements in the Fisher information matrix only for these generators (as shown in the top right of Fig. \ref{fig:Fisher_setting}). In the second scenario we measure momentum quadratures $p_1$ and $p_2$ (see the left bottom scheme of Fig. \ref{fig:Fisher_setting}). In this case, only generators $\hat q_1$ and $\hat q_2$ displace the distribution in the $(p_1, p_2)$-setting resulting in a different Fisher information matrix. We thus obtain a set of Fisher information matrices $F(\hat{\rho},\mathcal{A}_{N}, \phi)$ depending on the state $\hat \rho $, on the set of generators $\mathcal{A}_N$ and on the measurement setting parametrized by $\phi$.

In all possible measurement settings, the classical Fisher information matrix is a lower bound on the quantum Fisher information matrix, \ie $F(\hat{\rho}(\boldsymbol{\vartheta}),\mathcal{A}_{N}, \phi) \preccurlyeq Q(\hat{\rho}(\boldsymbol{\vartheta}),\mathcal{A}_{N})$, leading to the following measurement-setting-dependent entanglement witness
\begin{equation}\label{eq:weaker_entanglement_witness}
    E_{\rho(\boldsymbol{\vartheta}),\text{hom}}^{\mathcal A_N}(\phi)\equiv \lambda_{\rm \scriptscriptstyle max}\left(F(\hat \rho,\mathcal A_N, \phi)-4 \Gamma_{\Pi(\hat \rho(\boldsymbol{\vartheta}))}^{\mathcal A_N}\right).
\end{equation}
Here we remark the dependence on the mode basis $\boldsymbol{\vartheta}$ in which we probe the entanglement of the state $\hat \rho$. The objective is to then lift the criteria to probe in different basis, in the same way as we did for the quantum Fisher information matrix. This witness is a lower bound of the witness given in equation \eqref{eq:entanglement_witness_matrix_form}
\begin{equation*}
    E_{\rho(\boldsymbol{\vartheta}),\text{hom}}^{\mathcal A_N}(\phi)\leq E_{\rho(\boldsymbol{\vartheta})}^{\mathcal A_N}(\phi).
\end{equation*}
We can then transform the classical Fisher information measured in a mode basis $\boldsymbol{\vartheta}_{0}$, to any other basis $\boldsymbol{\vartheta}$ using the same transformation as for the QFI, in equation \eqref{eq:transform_QFI}.
The resulting matrix, though not necessarily a Fisher information matrix corresponding to a specific measurement, is still a lower bound of the quantum information matrix in the new basis, \ie 
\begin{equation*}
   U(\boldsymbol{\vartheta}) F(\hat{\rho}(\boldsymbol{\vartheta}_0),\mathcal{A}_{N}, \phi) U(\boldsymbol{\vartheta})^{\top} \preccurlyeq U(\boldsymbol{\vartheta}) Q(\hat{\rho}(\boldsymbol{\vartheta}_0),\mathcal{A}_{N})U(\boldsymbol{\vartheta})^{\top}.
\end{equation*}
This allows us to use the classical Fisher information matrix obtained in a specific basis $\boldsymbol{\vartheta}_0$ to probe any other basis $\boldsymbol{\vartheta}$. It is worth noticing that once we probe in this way a new basis, the witness we obtain is dependent on $\boldsymbol{\vartheta}_0$, and is not the same as that in Eq. \eqref{eq:weaker_entanglement_witness}. We define it as 
\begin{equation}\label{eq:initial_base_dependent_witness}
\begin{split}
    & E_{\hat \rho,\text{hom}}^{\mathcal A_{N}}(\phi,\boldsymbol{\vartheta}|\boldsymbol{\vartheta}_0)\equiv \\
    & \lambda_{\rm \scriptscriptstyle max}\left(U(\boldsymbol{\vartheta})F(\hat \rho,\mathcal A_N, \phi)U(\boldsymbol{\vartheta})^{\top}-4 \left(U(\boldsymbol{\vartheta})\Gamma^{\mathcal A_N}U(\boldsymbol{\vartheta})^{\top}\right)_{\Pi(\hat \rho)}\right).
    \end{split}
\end{equation}

Using this witness, we can, similarly as in Eq.~\eqref{eq:PS-metrological-witness}, define a (measurement setting dependent) mode-intrinsic entanglement witness given by 
\begin{equation*}\label{eq:weak_witness_mode_intrinsic_entanglement}
    \underset{\boldsymbol{\vartheta}}{\text{min}} \,E_{\hat \rho,\rm  hom}^{\mathcal A_{\scriptscriptstyle N}}(\phi,\boldsymbol{\vartheta}|\boldsymbol{\vartheta}_0) >0.
\end{equation*}
We can create stronger versions of this witness by considering for each mode basis $\boldsymbol{\vartheta}$ the best value of the witness $E_{\hat \rho,\text{hom}}^{\mathcal A_N}(\phi^{(i)},\boldsymbol{\vartheta}|\boldsymbol{\vartheta}_0)$ for different homodyne settings $\{\phi^{(1)}, \ldots,\phi^{(n)}\}$, and then minimizing over the mode-basis parameter space $\boldsymbol{\vartheta}$:
\begin{equation}\label{eq:weak_witness_mode_intrinsic_entanglement_n_settings}
    \mathcal{W}_{\text{hom}}(\hat \rho,\mathcal A_N|\boldsymbol{\vartheta}_0) \equiv \underset{\boldsymbol{\vartheta}}{\text{min}} \,\underset{\boldsymbol{\{\phi^{(1)}, \ldots,\phi^{(n)}\}}}{\text{max}} \,E_{\hat \rho,\text{hom}}^{\mathcal A_{\scriptscriptstyle N}}(\phi^{(i)},\boldsymbol{\vartheta}|\boldsymbol{\vartheta}_0). 
\end{equation}
Usually we will consider $\boldsymbol{\vartheta}_0=\mathbf{0}$, which corresponds to measure directly in the basis in which the state is prepared. As it turns out, this seems to be the best choice for the kind of states we consider, \ie states obtained by photon subtraction in a superposition of two independently squeezed modes. For other kind of states, an exploration of different initial basis for the measure is advised to potentially increase the performance of the witness.\par  
In what follows we will apply this procedure on two-mode two-photon subtracted states that we have considered in previous sections. In all cases the plots were constructed using the Fisher information matrices corresponding to the measurement settings $\{q_1,q_2\}$ and $\{p_1,p_2\}$ (as in Fig. \ref{fig:Fisher_setting}), which are the maximally informative homodyne settings for the states that we consider. We apply the criteria on both of these Fisher information matrices as explained above and for each mode basis $\boldsymbol{\vartheta} = (\theta, \varphi)$  we pick the highest value between the two of them. This precludes us from doing optimization for the determination of the passive separability, but in general, at order two of the generators and for a low number of modes it is still a fast procedure. \par

 We can observe in Fig.~\ref{fig:class_FI_Criteria} an analysis of the evaluation of the criteria at second order of the generators, on two photon subtracted states under the effect of losses. In particular, we consider, as in Fig.~\ref{fig:lossy_2photonSubState_BothOrder_r1_1Pt5dB} photon subtraction from two squeezed modes with squeezing levels $s_1=1.5$ dB and $s_2=-2.6$ dB, for which some loss resilience is expected \cite{David_2023}. In the top row of Fig.~\ref{fig:class_FI_Criteria} we consider the case in which the two photons are subtracted in different modes parametrized by $\Theta_1=\pi/4$ and $\Theta_2=-\pi/4$, respectively. We can observe that the witness is positive in all bases for losses lower than $10\%$. In the bottom row we show the behavior of the witness when both photons are subtracted in the same mode, parametrized by $\Theta=\pi/4$. In this case the witness becomes negative around $8\%$ loss rate. In both cases the resilience to losses of the mode-intrinsic entanglement witness is quite low. Nevertheless, it is worth analyzing the values of the witness at different mode basis in the presence of losses. The second column of Fig.~\ref{fig:class_FI_Criteria} shows the contour plot of the entanglement witness over a grid of relevant mode basis change at a level of losses of $5\%$. In this case both plots have positive values in all mode basis, with very large values in some regions. On the other hand in the third column we show the behavior for a level of losses of $20\%$. At this point we can observe that both plots have regions in which the witness is non-positive, while still having positive values in a large set of basis. Even though we are not witnessing mode-intrinsic entanglement in this case, which is the central concern of this paper, these plots show the potential of the method we have developed for the less specific task of entanglement detection in different mode bases. We can think of it as a way to unveil hidden entanglement measuring in only one basis. In the cases that we considered the highest values of the witness were actually obtained for basis in which we did not perform any measurement on the state.  \par 
The results shown in Fig.\ref{fig:class_FI_Criteria} show that this method has some potential experimental utility, nevertheless, the fact that this is done at second order in the generators makes it far from the reach of current experimental capabilities. In the next subsection we show a more experimentally friendly approach, by evaluating the criteria at first order in the same kind of states. We will compute all the quantities involved in the criteria from homodyne data simulated to resemble what we would obtain from an actual experimental realization. In the suplementary material \cite{supp_mat} we perform a detailed comparison, for the same states analysed in Fig.\ref{fig:class_FI_Criteria}, between the results that are obtained using the method we have developed, and those obtained using a Gaussian witness, namely the Duan et \textit{al.} criterion \cite{Duan2000}.

\subsection{Implementation of simulated homodyne data}

\begin{figure*}
    \centering
    \includegraphics[width=\textwidth]{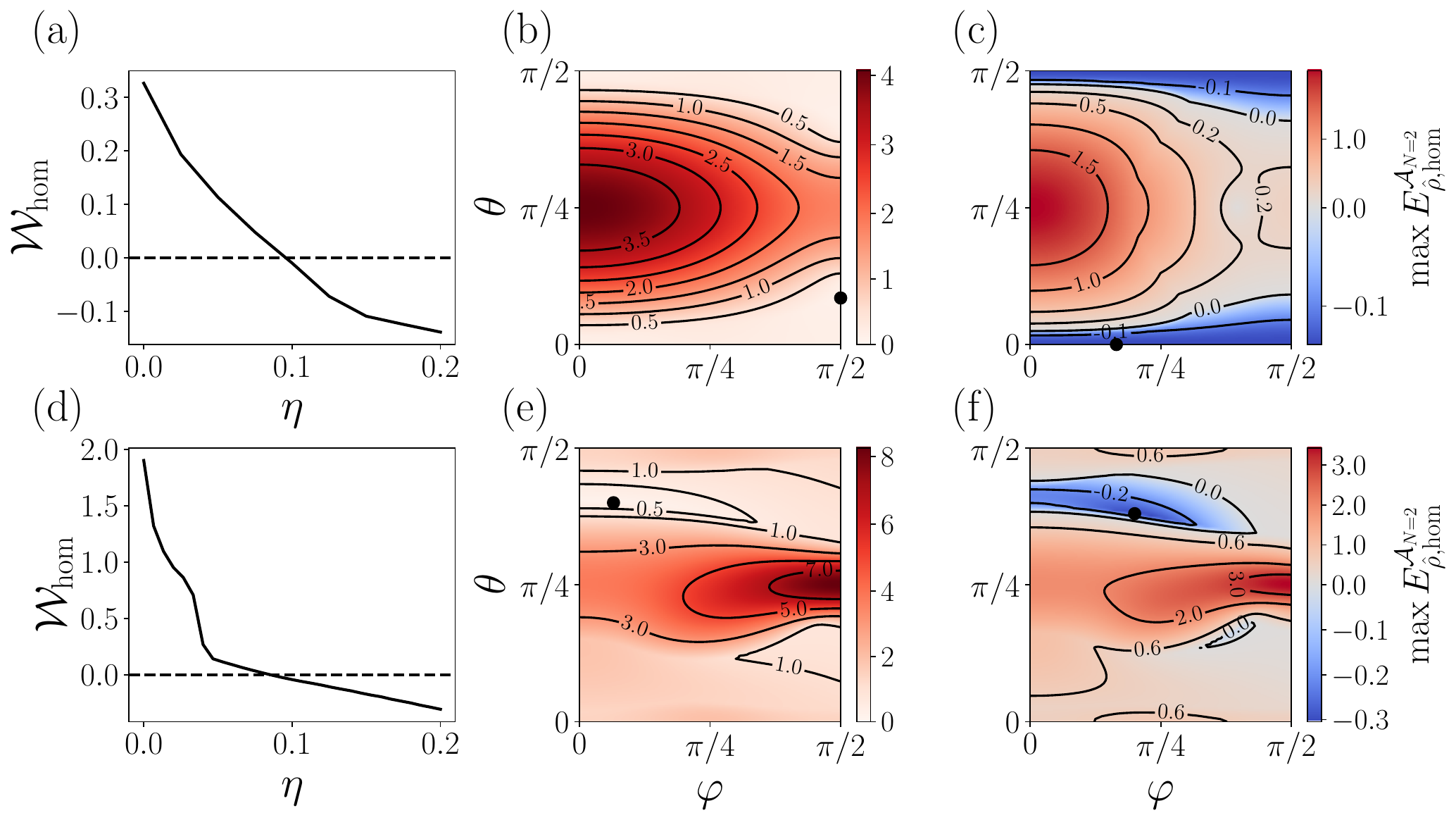}
    \caption{Behavior of the metrological witness for two-mode two-photon subtracted states, under the effect of losses. We consider that the two photons are subtracted from two single mode squeezed states with squeezing levels $r_1=1.5$ dB and $r_2=-2.6$ dB. In the top row we consider the two photon subtractions to happen in complementary modes ($\Theta_1=\pi/4$, $\Theta_2=-\pi/4$), while in the bottom row we consider the subtractions to happen in the same mode, given by $\Theta_1 = \Theta_2 = \pi/4$.
    The first column shows the metrological witness derived in Eq. \eqref{eq:weak_witness_mode_intrinsic_entanglement_n_settings} $\mathcal{W}_{\text{hom}}(\hat \rho,\mathcal A_{N=2}|\boldsymbol{\vartheta}_0)$.
    The second and third column show the behavior of the mode-specific entanglement witness $\underset{\boldsymbol{\{\phi^{(1)},\phi^{(2)}\}}}{\text{max}} \,E_{\hat \rho,\text{hom}}^{\mathcal A_{\scriptscriptstyle N = 2}}(\phi^{(i)},\boldsymbol{\vartheta}|\boldsymbol{\vartheta}_0)$ for the measurement settings $\phi^{(1)} = \{q_1, q_2\}$ and $\phi^{(2)} = \{p_1, p_2\}$ and for all relevant mode basis at levels of losses of $5\%$ and $20\%$ respectively. }
    \label{fig:class_FI_Criteria}
\end{figure*}

\begin{figure*}[h!]
    \centering
    \includegraphics[width=\textwidth]{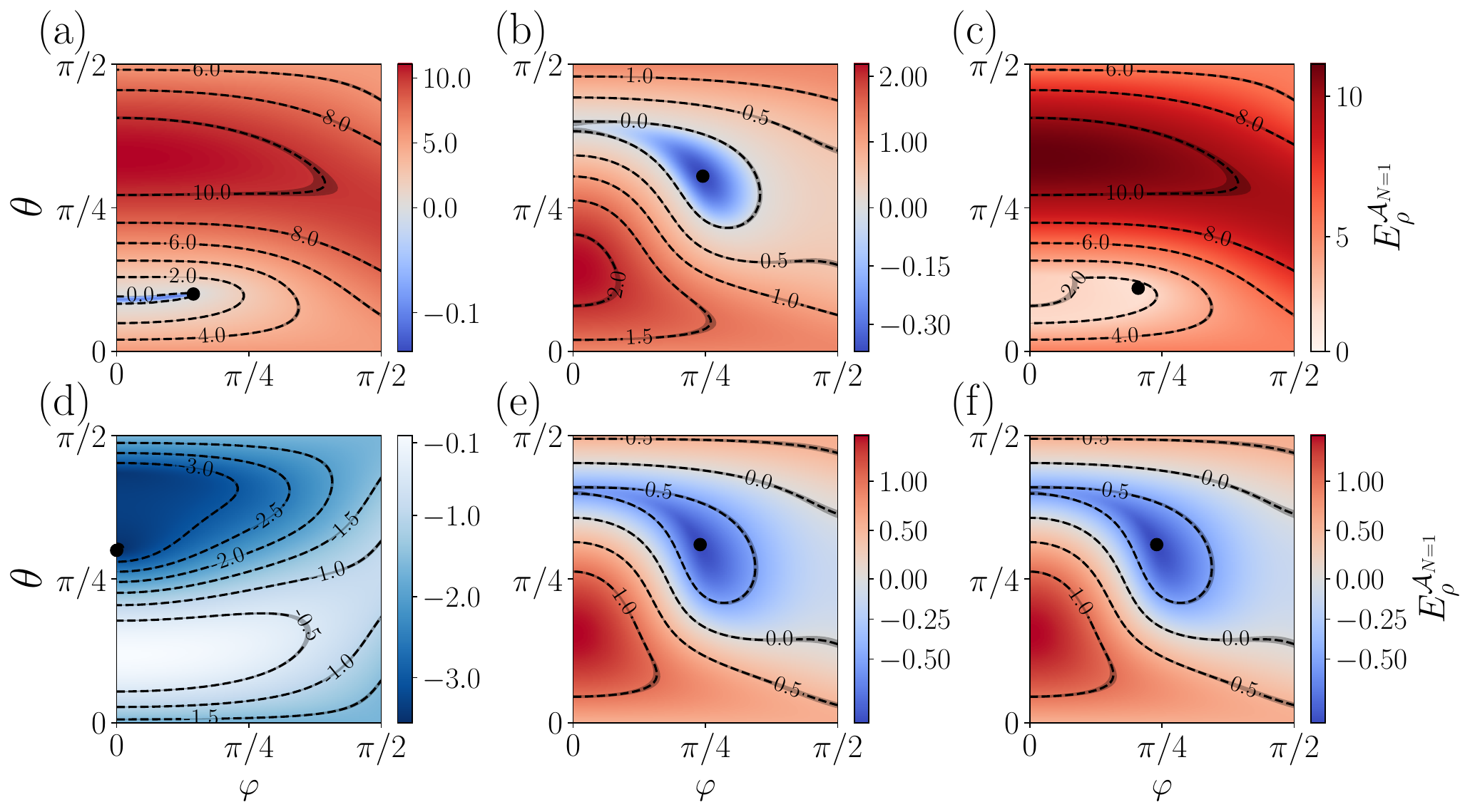}
    \caption{Simulation of an experimental evaluation of the witness \eqref{eq:weak_witness_mode_intrinsic_entanglement_n_settings} averaged over 100 samples, each of them containing 1000000 measurement points obtained from homodyne detection. We consider here a two-photon subtracted state with squeezing levels $r_1=1.5$ dB and $r_2=-2.6$ dB and subtractions in the same mode $\Theta_1 = \Theta_2 = \pi/4$. Evaluation of the witness $E_{\rho}^{\mathcal{A}_{\scriptscriptstyle N = 1}}$ in the measurement basis (a) $(q_1, q_2)$ [see Fig.~\ref{fig:Fisher_setting} (a)], (b) $(p_1, p_2)$ [see Fig.~\ref{fig:Fisher_setting} (d)]. (c) maximal value between both measurement settings $(q_1, q_2)$ and $(p_1, p_2)$ for each mode basis change $(\theta, \varphi)$ of the unitary transformation $U(\boldsymbol{\vartheta})$. The top row corresponds to zero loss, while the bottom row [plots (d), (e), (f)] are obtained from the same measurement settings as the top row but for 10\% of losses. The grey areas shown for each contour line correspond to the uncertainties resulting from the 100 samples.     }
    \label{fig:sampling}
\end{figure*}

We consider the simulation of an experimental evaluation of the witness \eqref{eq:weak_witness_mode_intrinsic_entanglement_n_settings}. We should keep in mind that this witness is a lower bound of the one in equation \eqref{eq:PS-metrological-witness}. We consider in particular the evaluation of the witness relying on two homodyne settings $\phi_1=\{0,0\}$, and $\phi_2=\{\pi/2,\pi/2\}$, which correspond to Alice and Bob measuring simulateously in the same quadratures $\hat q_A, \hat q_B$ and $\hat p_A, \hat p_B$ respectively. Our goal is to explore the relevance of the protocol for currently feasible experiments, for that reason we restrict ourselves to the evaluation of the criteria for first order generators, \ie $\mathcal A^{(1)}=\{\hat q_A,\hat q_B, \hat p_A,\hat p_B\}$, whose effect can be accounted for in post-processing of the measurement data. \par 

To evaluate the criteria we sample $1 0^6$ points from the probability marginals associated to the different measurement settings $\phi^{(i)} =\{\phi_1^{(i)},\phi_2^{(i)}\}$, to obtain a histogram of relative frequencies $\mathcal{F}(\xi_{\phi_1^{(i)}},\xi_{\phi_2^{(i)}})$ that in the limit of a large number of samples should converge to the exact distribution $p(\xi_{\phi_1^{(i)}},\xi_{\phi_2^{(i)}})$, obtained through Eq. \eqref{eq:marginal_prob_distrib_parametrized}. Once we have all the frequency histograms we use them to compute the classical Fisher information matrices associated to each of them, which can be done through the use of the Hellinger distance \cite{Braunstein1994}, defined as 
\begin{equation}\label{eq:Hellinger_distance}
\begin{split}
  d_H^2(\boldsymbol{\varphi})\equiv  \int_{\xi_{\phi_1^{(i)}},\xi_{\phi_2^{(i)}}} & \left( \sqrt{p(\xi_{\phi_1^{(i)}}, \xi_{\phi_2^{(i)}}|\boldsymbol{\kappa})} \right. \\
  &  \quad \quad  \left. -\sqrt{p(\xi_{\phi_1^{(i)}}, \xi_{\phi_2^{(i)}}|0)}\right)^2  d\xi_{\phi_1^{(i)}} d\xi_{\phi_2^{(i)}},  
\end{split}
\end{equation}
where $p(\xi_{\phi_1}, \xi_{\phi_2}|\boldsymbol{\kappa})$ is the probability distribution after the implementation of the parameter $\boldsymbol{\kappa}$, which in our case corresponds to a displacement, that can be trivially evaluated following \eqref{eq:effect_displacements}. In the computation of the Hellinger distance we replace the exact probability distribution by the corresponding histograms and consequently the integral is approximated by a discrete sum. For small values of the parameter the Hellinger distance can be approximated by the quadratic form 
\begin{equation}
    d^2_{H}(\boldsymbol{\kappa})\approx  \boldsymbol{\kappa}^T \frac{F}{8} \boldsymbol{\kappa} +\mathcal O(|\boldsymbol{\kappa}|^3), 
\end{equation}
where $F$ is the Fisher information matrix associated to the histogram. \par 

In Fig.\ref{fig:sampling} we show the behavior of the entanglement witness on different bases for the same states considered in the bottom row of Fig.\ref{fig:class_FI_Criteria}. The quantities in each plot were computed  over 100 realizations of the sampling process, which allowed to estimate the uncertainties on the values of the witness. These uncertainties are represented through the gray areas surrounding each contour. The top row of Fig.\ref{fig:sampling} corresponds to the pure state case. In the bottom row we consider the state after a lossy channel with $10\%$ loss rate. The first column shows, for both instances, the evaluation of the witness \eqref{eq:weaker_entanglement_witness} from measuring simultaneously the momentum quadratures in both modes. In the second column we show the evaluation from the measurement of the position quadratures. In the lossy case only the position quadrature based witness gives positive results, which is related to the inhomogeneous distribution of squeezing in the Gaussian state from which the photons were subtracted. While we are not able to detect mode-intrinsic entanglement at this level of losses, these results show the potential of the method we have developed to detect entanglement in a wide range of bases, in an experimentally feasible way and from a rather restricted set of measurement settings. This last remark also emphasizes and strengthens the ability to detect entanglement features from a limited set of probability marginals, thereby justifying the high effectiveness of neural network methods that rely on the same type of correlation patterns to identify entanglement \cite{Xiaoting_2024}.


\section{Conclusion}

In this paper, we derive a witness able to detect a specific form of entanglement only exhibited by non-Gaussian states and considered as an important resource to reach a quantum computational advange in a family of sampling protocols in the context of CV systems \cite{Ulysse_Mattia_2022}: \textit{mode-intrinsic entanglement}. Contrary to Gaussian states for which a mode basis where the state is separable always exists, entanglement of certain non-Gaussian states cannot be destroyed by linear passive operations and is present in all mode bases. We refer to this type of entanglement as mode-intrinsic entanglement. As we demonstrate in this work, our mode-intrinsic entanglement witness based on metrological tools, and more specifically on the quantum Fisher information, proves to be efficient when applied to experimentally relevant states such as multimode photon subtracted states. 

Moreover, we also underline in the last part of the paper the capability of the witness to be experimentally accessible. First, it only requires measurements in one mode basis. Indeed, we show that once the witness has been measured in one basis, one can check entanglement in any other basis in post-processing. Second, while the witness relies on the quantum Fisher information, a quantity very difficult -- if not impossible, to access experimentally, one can use a lower bound of the witness through the classical Fisher information. As we explain in Sec.~\ref{sec:relaxation_criteria}, this witness can be evaluated through different homodyne measurement settings and give rather good results. We highlight and check its experimental applicability with simulations of experimental data obtained from probability marginals.

Overall, what we propose is a new and promising tool to explore the specificity of quantum correlations exhibited by non-Gaussian states. Even if the witness is sensitive to losses, thus limiting its capability to certify mode-intrinsic entanglement, it works in a wide range of mode bases and in some cases surpasses the efficiency of Gaussian witnesses. 

In future studies it would be nice to find methods to mitigate the effect of losses; one could for example complement our witness with other entanglement criteria in specific bases where it fails to detect entanglement. We would also like to explore in more detail the results that we obtained for multimode cluster states, in particular how non-Gaussian operations like photon subtraction modify the entanglement properties of the cluster. Another aspect would be to better understand the interplay between entanglement and the metrological advantage that non-Gaussian states offer; or, in other words, to which extent states that strongly violate the metrological witness can be useful for some applications in metrology.

\begin{acknowledgments}
We thank M. Frigerio for inspiring discussions. We acknowledge financial support from the ANR JCJC project NoRdiC (ANR-21-CE47-0005), the Plan France 2030 through the project OQuLus (ANR-22-PETQ-0013), the HORIZON-EIC-2022- PATHFINDERCHALLENGES-01 programme under Grant Agreement Number 101114899 (Veriqub), and the QuantERA II project SPARQL that has received funding from the European Union’s Horizon 2020 research and innovation programme under Grant Agreement No 101017733
\end{acknowledgments}

\bibliography{biblio}

\end{document}